\documentclass[12pt]{article}

\usepackage{slashed}
\usepackage{color, verbatim}
\usepackage{latexsym}
\usepackage{amsmath}
\usepackage{graphicx}
\usepackage{arydshln}

\usepackage{cite}
\usepackage{hyperref}

\usepackage{amssymb} 
\usepackage{graphicx}
\usepackage{bm}
\usepackage{hyperref}
\usepackage{adjustbox}
\usepackage{booktabs}
\usepackage{multirow}
\usepackage{rotating}
\usepackage{multirow, array}
\usepackage{float}

\makeatletter

\@addtoreset{equation}{section}
\makeatother

\newcommand{\ii}{\ensuremath{\mathrm{i}}}

\makeatletter
\g@addto@macro\bfseries{\boldmath}
\makeatother

\begin{document}
\thispagestyle{empty}

\vspace{0.8cm}

\begin{center}

\begin{center}

{\Large 
Following the muon track of hierarchical sectors at LHCb}

\end{center}

\vspace{0.8cm}

\textbf{Xabier Cid Vidal$^{\,a}$, Titus Momb\"acher$^{\,a}$, Maria Ramos$^{\,b}$, Emilio Xosé Rodríguez Fernández$^{\,a}$}\\

\vspace{1.cm}

{\em {$^a$IGFAE,Universidade de Santiago de Compostela, Santiago de Compostela, Spain}}\\[0.2cm]

\it {$^b$Departamento de Fisica Teorica, Universidad Autonoma de Madrid,
and IFT-UAM/CSIC, Cantoblanco, 28049, Madrid, Spain}\\[0.2cm]
\it{E-mail:} \href{mailto:xabier.cid.vidal@cern.ch}{xabier.cid.vidal@cern.ch}, \href{mailto:titus.mombacher@cern.ch}{titus.mombacher@cern.ch}, \href{mailto:maria.pestanadaluz@uam.es}{maria.pestanadaluz@uam.es}, \href{mailto:emilio.xose.rodriguez.fernandez@cern.ch}{emilio.xose.rodriguez.fernandez@cern.ch}

\vspace{0.5cm}

\end{center}

\begin{abstract}
This paper reports a study of the experimental signatures of hierarchical sectors beyond the Standard Model characterized by a flavour-violating heavy vector and a set of light pseudo-Goldstone bosons $a_{1,2}$, spanning a large range of lifetimes. The non-minimal scalar spectrum triggers novel $B$ decays into multiple leptons that would have escaped the reach of current searches. Novel displaced vertex analyses at LHCb are therefore discussed to probe the hierarchical new physics, extending the scope of the tracking system of the detector. Additionally, the reach of the proposed CODEX-b experiment is studied. By exploiting the use of tracks only reconstructed in subsystems of the detector at the high-level LHCb trigger, ${\mathcal{B}(B_s^0\to a_1 a_2) < 10^{-8}}$ and ${\mathcal{B}(B^+\to K^+ a_1 a_2) < 10^{-9}}$ could be reached in the muon channel across seven orders of magnitude in the lifetime of the Goldstone bosons. Correspondingly, heavy-light particle couplings of order $\lesssim 1$ could be tested, potentially ruling out composite Higgs scenarios where the heavy and light sectors couple strongly.

\end{abstract}

\newpage

\tableofcontents

\newpage

\section{Introduction}

In spite of the strong experimental and theoretical reasons to expect departures from the Standard Model (SM), no direct signal of New Physics (NP) has been observed to date. Such paradigm admits two plausible interpretations: either the NP is too heavy, leading to signatures that can be studied in the context of the SM Effective Field Theory (SMEFT)~\cite{Brivio:2017vri} from a model independent perspective; or it is very feebly interacting, in which case our low-energy theory, the SM itself, must be extended with new light degrees of freedom (singlets being prominent candidates due to their elusive nature).

The two scenarios above are not exclusive if the new physics is hierarchical. This is a generic feature of ultra-violet (UV) complete models with a symmetry that is spontaneously broken at low energy, leading to the presence of both heavy and light fields -- the pseudo Nambu-Goldstone bosons (pNGBs) -- in the spectrum. %
Furthermore, combined solutions to the strongest anomalies observed in data might actually require a hierarchical spectrum: while heavy flavour violating vector bosons ($V$) are promising candidates to explain the central-bin anomalies observed in lepton flavour universality ratios~\cite{Buras:2013qja,Crivellin:2015xaa,Alguero:2022est}, lighter singlets ($a$) can more easily accommodate the anomalies in the low-$q^2$ bin~\cite{Sala:2017ihs,Altmannshofer:2017bsz}, as well as the tensions observed in the magnetic moment of leptons~\cite{Bauer:2017nlg,Liu:2018xkx,Buen-Abad:2021fwq}. These scenarios motivate the study of the experimental consequences of hierarchical couplings, namely in the $\{V,a\}$ framework. 

\begin{sloppypar}
At low energies, $V$ would trigger rare $B$-meson decays into the pNGBs that can subsequently decay into muons. LHCb has studied such processes in the prompt case involving up to four muons, presenting the stringent limit on the branching fraction \mbox{$\mathcal{B}(B_s^0 \to \mu^+ \mu^- \mu^+ \mu^-) < 8.6\times 10^{-10}$} at $95\,\%$ confidence level~\cite{LHCb:2021iwr}.
Furthermore, a limit was also determined on \mbox{$\mathcal{B}(B_s^0 \to a(\mu^+ \mu^-) a(\mu^+ \mu^-)) < 5.8\times 10^{-10}$} assuming a prompt 1\,GeV scalar, as proposed in Ref.~\cite{Chala:2019vzu}. Note that other studies have addressed the sensitivity of the LHCb experiment (or other $B$-factories) to similar or related theoretical setups \cite{Batell:2009jf,Borsato:2021aum,BuarqueFranzosi:2021kky,CidVidal:2018blh}.
Interestingly, when the scalar sector is non-minimal, final states with a larger number of muons can be expected since the decay of the heaviest $a_2\to a_1 a_1$ is typically more common than $a_2 \to \mu^+\mu^-$. To probe this scenario, novel analyses were proposed in Ref.~\cite{Blance:2019ixw} focusing on final states with six prompt muons.
\end{sloppypar}

Such high-multiplicity $B$-signatures arise naturally in non-minimal composite Higgs models (CHM)~\cite{Gripaios:2009pe,Sanz:2015sua,Chala:2016ykx} where the couplings between the light pNGBs and heavy composite vectors are predicted to be $\mathcal{O}(1)$ due to strong dynamics. Notwithstanding, the exotic scalar interactions are dependent on the quantum numbers of the heavy resonances that trigger their masses, in the spirit of {\it partial compositeness}~\cite{KAPLAN1991259,Panico:2015jxa}. Consequently, the scalar lifetimes can only be defined in concrete models and can {\it a priori} span several orders of magnitude. In fact, that would be the case if the new composite sector resembles at all that of the SM, where the lifetimes of the Goldstone particles are several orders of magnitude apart~\cite{Alimena:2019zri}.

Therefore, in this work, the assumption that the exotic sector is short-lived is relaxed and instead the prospects to detect the light scalars at LHCb across a large range of lifetimes are investigated. 
From an experimental point of view, to enhance the sensitivity for large displacements, novel analyses are proposed that do not require the tracking stations closest to the $pp$-collision vertex to reconstruct the longest-lived scalar in the discussed framework but instead rely on tracks only reconstructed in subsystems of the detector. To the best of the authors' knowledge, no search at LHCb up to date has considered such tracks because their reconstruction had not been included in the trigger sequence. This work therefore constitutes an important motivation for the inclusion of incomplete tracks at the high-level LHCb trigger and their usefulness for future displaced searches.
To investigate the sensitivity at extremely high lifetimes, the reach of the proposed CODEX-b experiment is also  analysed.

The paper is organized as follows. In Section~\ref{sec:L}, the relevant Lagrangian for this study is defined, as well as its constraints. 
In Section~\ref{ref:sec-limits}, the reach of LHCb with several track types is studied,  depending on the lifetime of the decaying particles. 
The expected backgrounds of the displaced analyses are discussed in detail.
The projected sensitivity for the Codex-b experiment is also presented. The interpretation of the upper limits within the composite Higgs scenario is discussed in Section~\ref{sec:model}. Finally, Section~\ref{sec:summary} is dedicated to the  conclusions. Two appendices are included to support the discussion in the previous sections. In Appendix~\ref{sec:embedding}, an explicit CHM is constructed, from which predictions to the parameters in the generic Lagrangian are obtained. In Appendix~\ref{sec:maps}, mass-dependent efficiency distributions are presented, which generalize the results in Section~\ref{ref:sec-limits}.

\section{Theoretical framework and constraints}~\label{sec:L}
The SM is considered to be extended with a flavour-violating heavy vector $V$ and two light scalars $a_{1,2}$ of different CP charge.
We assume that all new particles are singlets of the SM gauge group and that CP is a symmetry of the new interactions.
At low energies, the relevant Lagrangian of the beyond the SM (BSM) sector can be parameterized as~\cite{Blance:2019ixw}: 
\begin{equation}
L_{\rm eff} \supset \left[ g_{qa} (\overline{b_L} \gamma^\mu s_L)  (a_1 \overleftrightarrow{\partial_\mu} a_2) +  \text{i} g_1 y^\ell a_1 \overline{\ell} \gamma_5 \ell + g_2 y^\ell a_2 \overline{\ell} \ell +  {\rm h.c.}\right] - \frac{m_1^2}{2} a_1^2 - \frac{m_2^2}{2} a_2^2 - m_{12} a_2 a_1^2\,,
\label{eq:Lsetup}
\end{equation}
where $b$ and $s$ denote the SM bottom and strange left-handed quarks, $\ell$ a SM lepton and $y^\ell$ the corresponding SM Yukawa coupling.
We take $m_2 > m_1$ without loss of generality.  
The first term in the equation above results from the $V$ exchange at tree level, leading to
\begin{equation}
g_{qa} = \frac{g_{sb} g_{12}}{m_V^2}\,,  
\label{eq:gqa}
\end{equation}
with $m_V$, $g_{sb}$ and $g_{12}$ denoting, respectively, the vector mass, its coupling to the third generation quarks and to the exotic pNGBs. Note that the coupling in equation~\ref{eq:gqa} is not only suppressed by mass powers, but also by the CKM factors that relate $g_{sb}$ with the coupling $V \overline{q_L} q_L$ in the unbroken phase.

The muon channel is assumed to dominate the scalar decay width. This is satisfied in muonphilic scenarios, or in low-mass leptophilic scenarios if the scalars couple to leptons according to their masses. The latter scenario can arise if the scalar couplings to the SM fermions are protected by an approximate shift symmetry broken only by the mass terms in the Lagrangian, as commonly assumed in studies of axion-like particles connected to the strong CP problem~\cite{Rubakov:1997vp,Berezhiani:2000gh,Gaillard:2018xgk,Agrawal:2017ksf,Bauer:2021mvw}. Furthermore, this study explores the regime where ${g_1 \ll g_2 \sim \mathcal{O}(1)}$. In this case, the CP-even scalar with $m_2\sim 1$\,GeV can provide an explanation to the $(g-2)_\mu$ anomaly within $2\sigma$~\cite{Chala:2019vzu}, without the latter being spoiled by the contribution of the CP-odd one (that contributes to the anomaly with the wrong sign~\cite{Bauer:2021mvw}\footnote{An induced photon coupling at the loop level could change the sign of the axion-like particle contribution. Still, to be compatible with the muon anomaly, flavour non-universal $g_1$ couplings would be required.}).
Given this hierarchy between the couplings, the lightest scalar is expected to be long-lived, while the heaviest one will decay promptly into $a_1 a_1$ ($\mu^+\mu^-$) if $m_2 > 2 m_1$ ($m_2 < 2 m_1$). This holds provided that $m_{12}/m_2 \gtrsim g_2 y^\ell$. A possible CHM compatible with these assumptions is presented explicitly in Appendix~\ref{sec:embedding}.

The effective interactions trigger at low-energy $B_s^0 \to a_1 a_2$ decays leading to at least one pair of displaced muons in the detector.  
In the limit $m_1 = m_2$, the two scalars transform as a complex scalar field that couples to $\partial B$ through a Noether current. Consequently, the two-body decay width vanishes~\cite{Blance:2019ixw}. To probe the model in this limit, the additional channel $B^+ \to K^+ a_1 a_2$ is considered.

The Yukawa couplings in equation~\ref{eq:Lsetup} are subject to collider constraints. 
Previous searches from BaBar in $e^+e^- \to \mu^+ \mu^- X (\mu^+ \mu^-)$ final states do not however constrain these couplings due to the lepton mass suppression~\cite{Liu:2018xkx}. 
On the other hand, the recent CMS search~\cite{CMS:2021sch} for long-lived dimuon resonances with a dedicated dimuon trigger provides powerful constraints in the parameter space. Although a complete simulation is beyond the scope of this work, the search has been recast to extract the order of magnitude reach of the constraints on the proposed models. Among the most important selection cuts on events that pass the first level of the trigger, the presence of at least two muons with $p_T^\mu > 4$\,GeV and a large transverse replacement of the decay vertex with respect to the primary interaction vertex are required. The search by CMS is performed in categories of isolated and non-isolated muons. The recast is simplified by assuming the efficiencies of non-isolated muons (these are higher than the efficiencies when additionally isolation requirements are imposed). Thus the obtained limits from the recast are considered to be significantly optimistic. Up to lifetimes of around $10$\,ns, the recast imposes limits down to branching fractions of $\mathcal{B} (B_s^0 \to a_1 a_2)\sim \mathcal{O}(10^{-7})$, becoming one (two) orders of magnitude weaker for lifetimes of $10^2\,(10^3)$\, ns. Additional sensitivity might be achieved by analysing the parked $B$-physics data of the CMS experiment. However, as will become clear in Section~\ref{ref:sec-limits}, the sensitivity reach of the dedicated LHCb analyses proposed in this work could surpass these limits by orders of magnitude already at Run 3, except for $\tau \gtrsim 10$\,ns.

Similarly, the recent LHCb search for low-mass dimuon resonances~\cite{LHCb:2020ysn} is sensitive to our scenario, in the fiducial region requiring that the resonance decays displaced from the $pp$ collision. Furthermore, the analysis requires that the reconstructed particle has a $p_T$ within the range ${[2,10]\,{\rm GeV}}$; that the secondary vertex is transversely displaced from the primary one; as well as a minimal separation between the two muons of $3$\,mrad. A recast of this search has been performed by determining the efficiency of the fiducial selection cuts per energy bin for the signal modes studied in this work. The following branching fractions were found: $\mathcal{B} (B_s^0 \to a_1 a_2)\sim 8\times10^{-7} (4\times10^{-6})$ for $\tau_1 \sim 1\,(100)$\,ps, with the efficiency vanishing for larger lifetimes. (The values $m_1=1\,$GeV and $m_2=2.5\,$ GeV were used as benchmarks). 
Again, with a dedicated analysis as proposed in this work the sensitivity to these final states can be further improved by about two orders of magnitude.
 
Finally, future Belle-II prospects could compete with the searches for ${B^+\to K^+a_1a_2}$ at LHCb
due to the low backgrounds and the larger reconstruction efficiency of the displaced vertex~\cite{Filimonova:2019tuy}, or even the possibility to avoid the need of reconstructing one of the final state bosons, an approach which has the advantage of not having any limitation in terms of lifetime reach  \cite{Ferber:2022rsf}. Assuming $50\,{\rm ab}^{-1}$ for the Belle-II experiment, the corresponding prospects would be limited with respect to LHCb due to 
the large statistics expected to be accumulated by LHCb by Run 5 (rendering an about $6$ times larger data set). However, considering an upgrade under discussion collecting $250\,{\rm ab}^{-1}$ of data~\cite{Ishikawa:2020gpo}, the searches for the $B^+$ decays at the Belle-II experiment could be competitive.

\section{Displaced searches at LHCb and Codex-b}~\label{ref:sec-limits}

\begin{figure}[t]
    \centering
    \includegraphics[width=0.65\textwidth]{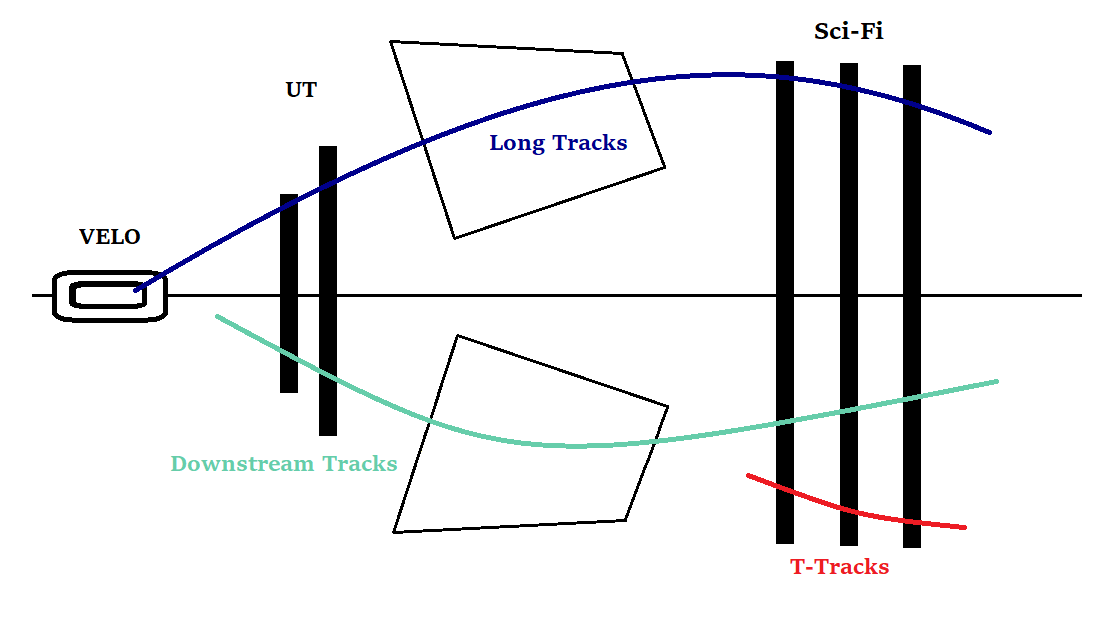}
    \caption{\it Schematic drawing of the LHCb detector with the different types of tracks that are considered in this analysis.}
    \label{fig:lhcb_tracks}
\end{figure}

The good muon identification capabilities for relatively soft momentum tracks and the precise vertex resolution make the LHCb experiment an excellent choice to look for displaced multi-muon signatures from $B$-decays.
In the LHCb detector, tracks are reconstructed with various subsystems: the Vertex Locator (VELO); tracking stations upstream of the magnet (UT), the Scintillating Fibers stations (SciFi) downstream of the magnet and the muon system.
While the best resolution can be reached with tracks that are reconstructed in all subsystems, a good quality can also be maintained for tracks that are reconstructed only in parts of the detector.
Neglecting subsystems close to the interaction point allows to reconstruct particles with very large lifetimes, that is precisely the goal of this study. Therefore, several track types are considered in this analysis as specified in Table~\ref{tab:tracktypes} and sketched in Figure~\ref{fig:lhcb_tracks}.

Long tracks leave hits in all tracking systems of the LHCb detector, particularly in the VELO detector, which is closest to the interaction point. Downstream tracks have associated hits in all tracking systems but the VELO detector. Additionally, T-tracks only have associated hits in the SciFi detector.
For extremely long-lived particles we investigate the sensitivity of the planned CODEX-b experiment~\cite{Aielli:2019ivi,Aielli:2022awh} and add an extra class of tracks which could be reconstructed in the CODEX-b detector. The CODEX-b experiment is a proposed off-axis experiment designed for long-lived particles produced at the same interaction point as the LHCb experiment, which is planned to start taking data during Run 4 of the LHC. As such it is also potentially sensitive to these signatures for very large scalar lifetimes.

Following the motivations outlined in the previous section, the following  exclusive decay modes are studied:
\begin{enumerate}
    \item $B^0_s\to a_1(a_2\to a_1a_1)\to3\mu^+3\mu^-$ and $B^+\to a_1(a_2\to a_1a_1)K^+\to3\mu^+3\mu^-K^+$\,,
    \item $B^0_s\to a_1a_2\to2\mu^+2\mu^-$ and $B^+\to a_1a_2 K^+\to2\mu^+2 \mu^-K^+$\,,
\end{enumerate}
that correspond to the main decay modes of the scalar particles in the regimes $m_2> 2m_1$ and $m_2< 2 m_1$, respectively. While $a_2$ is taken as a prompt resonance, a range of lifetimes for $a_1$ is considered.

As a reference channel, the decay $B^0_s\to a_1 a_1\to2\mu^+2\mu^-$ is used, whose sensitivity is taken from the recently published search~\cite{LHCb:2021iwr}.
In that analysis, the LHCb experiment looked for a peak at the $B^0_s$ mass in the invariant four-muon mass, requiring that the two dimuon pairs have a reconstructed mass compatible with each other within the mass resolution.

\begin{figure}[t]
    \centering 
    \includegraphics[scale=0.6]{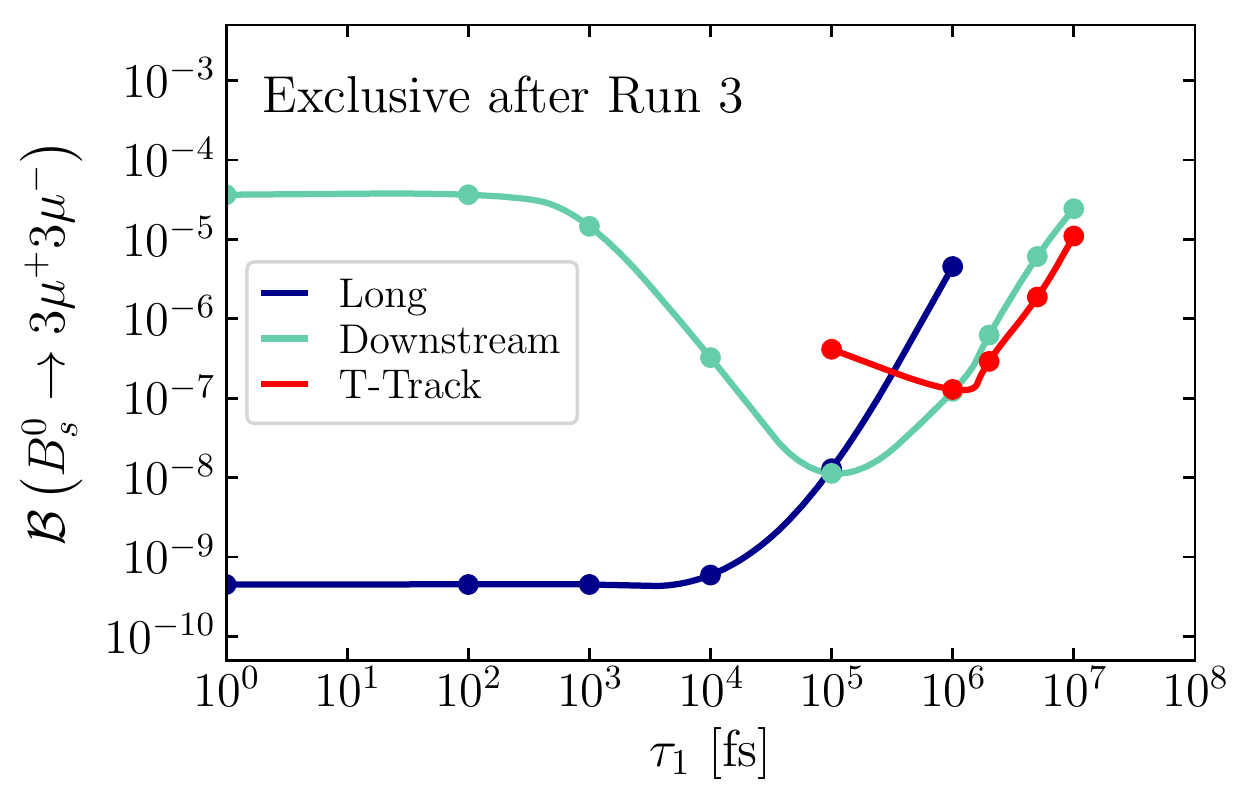}
    \includegraphics[scale=0.6]{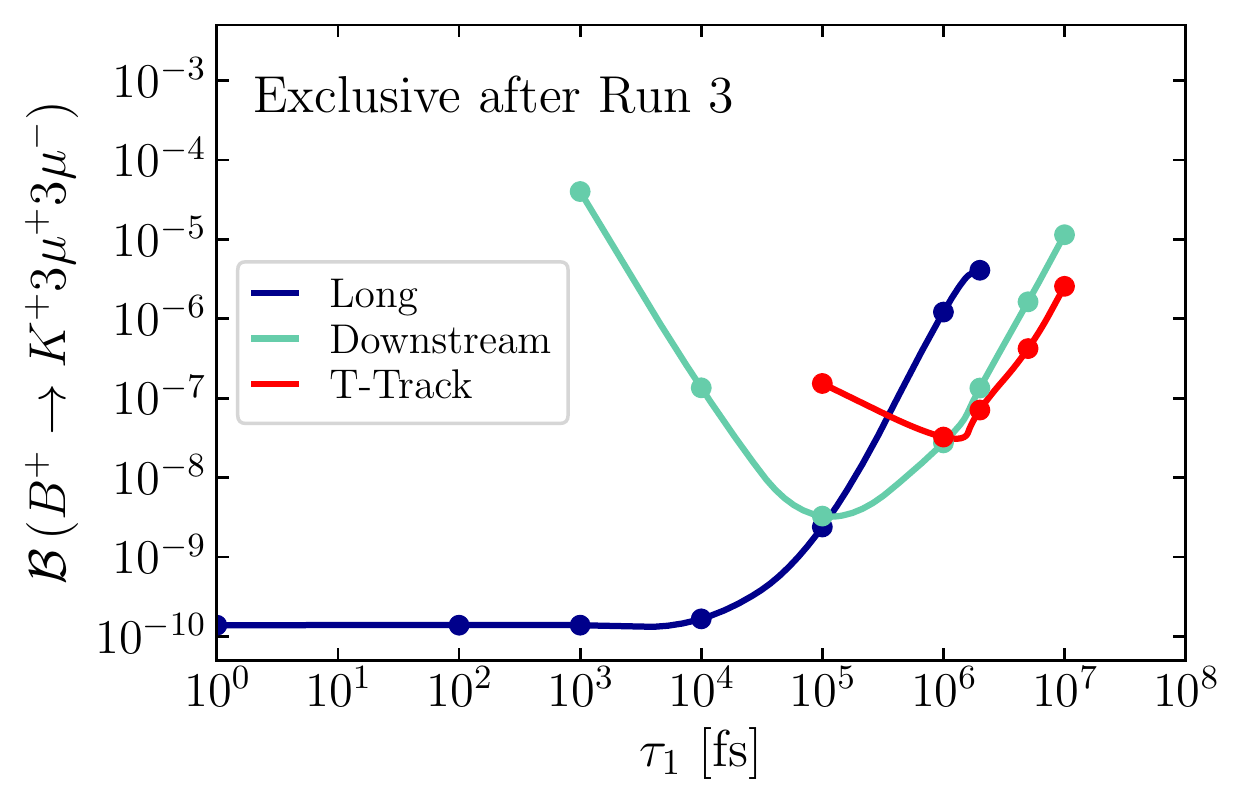}
        \includegraphics[scale=0.6]{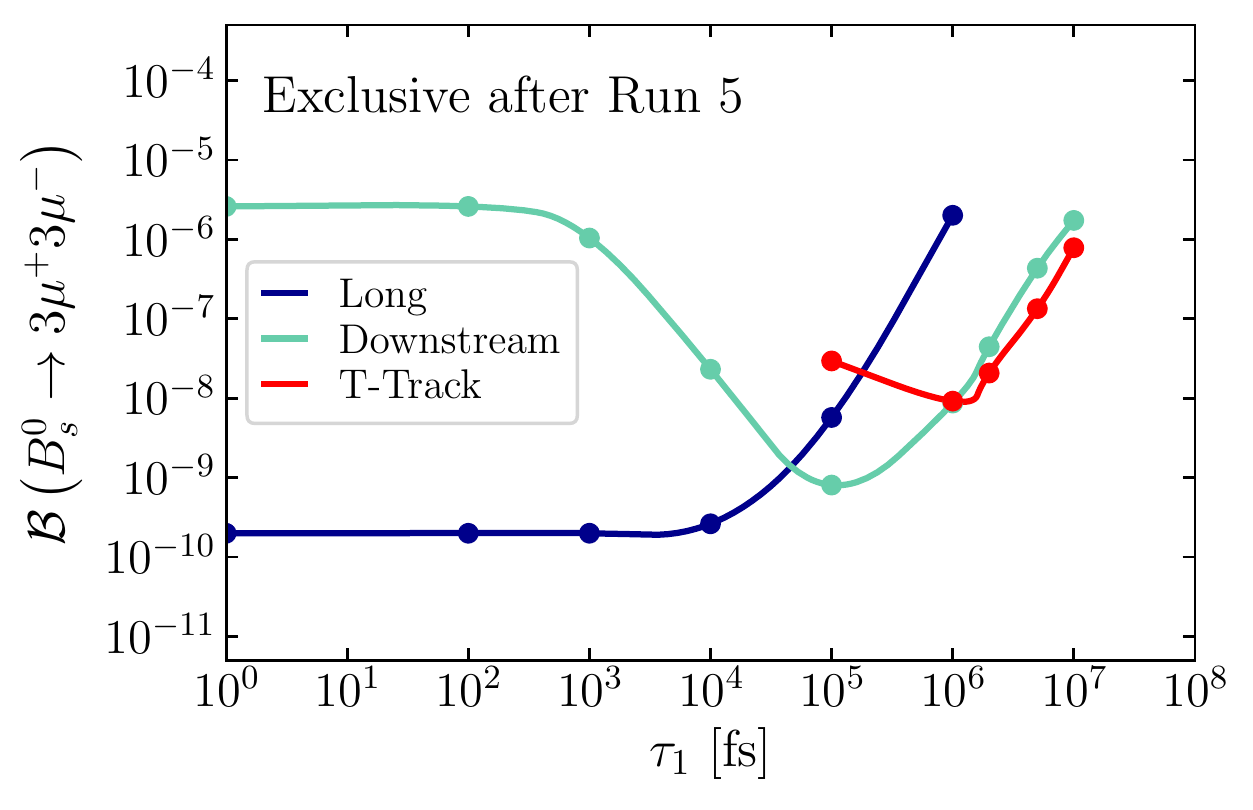}
    \includegraphics[scale=0.6]{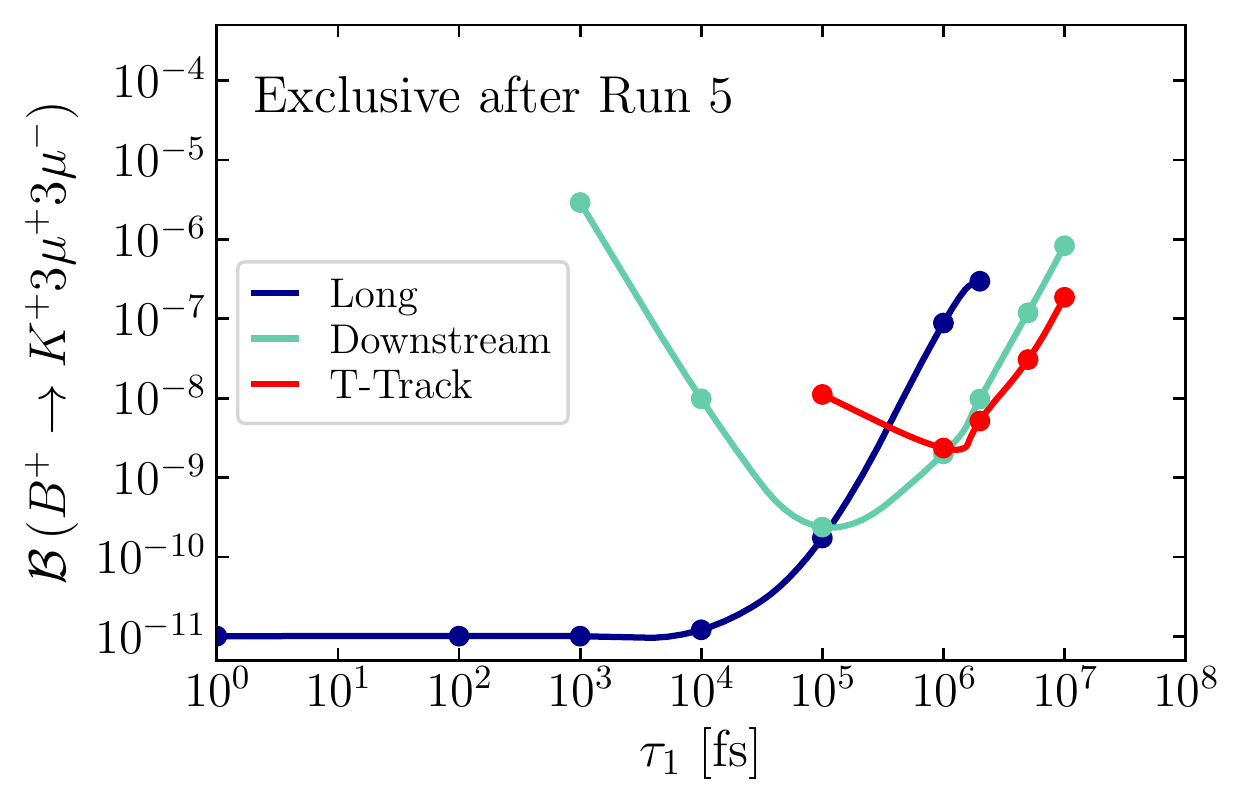}
    \caption{\it The expected upper limits on the $B^0_s\to 3\mu^+ 3\mu^-$ and $B^+\to K^+ 3\mu^+ 3\mu^-$ branching fractions that can be derived from the exclusive searches proposed in this work, for different lifetimes of the $a_1$ particle.  In the upper (bottom) panels, the detector configuration after Run 3 (Run 5) is assumed. The benchmark masses were fixed to $(m_1,m_2)=(1.0,2.5)$ GeV.}
    \label{fig:lim_6mu}
\end{figure}

\begin{figure}[t]
    \centering 
    \includegraphics[scale=0.6]{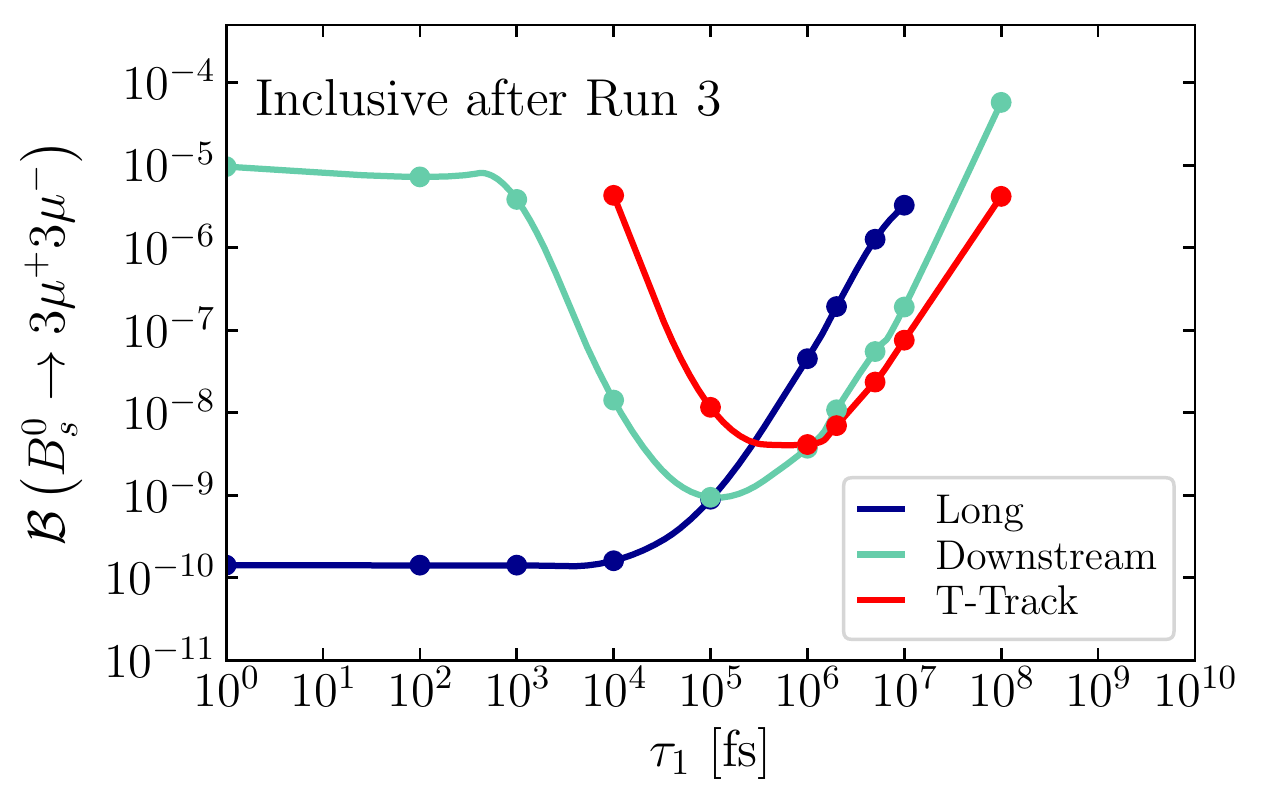}
    \includegraphics[scale=0.6]{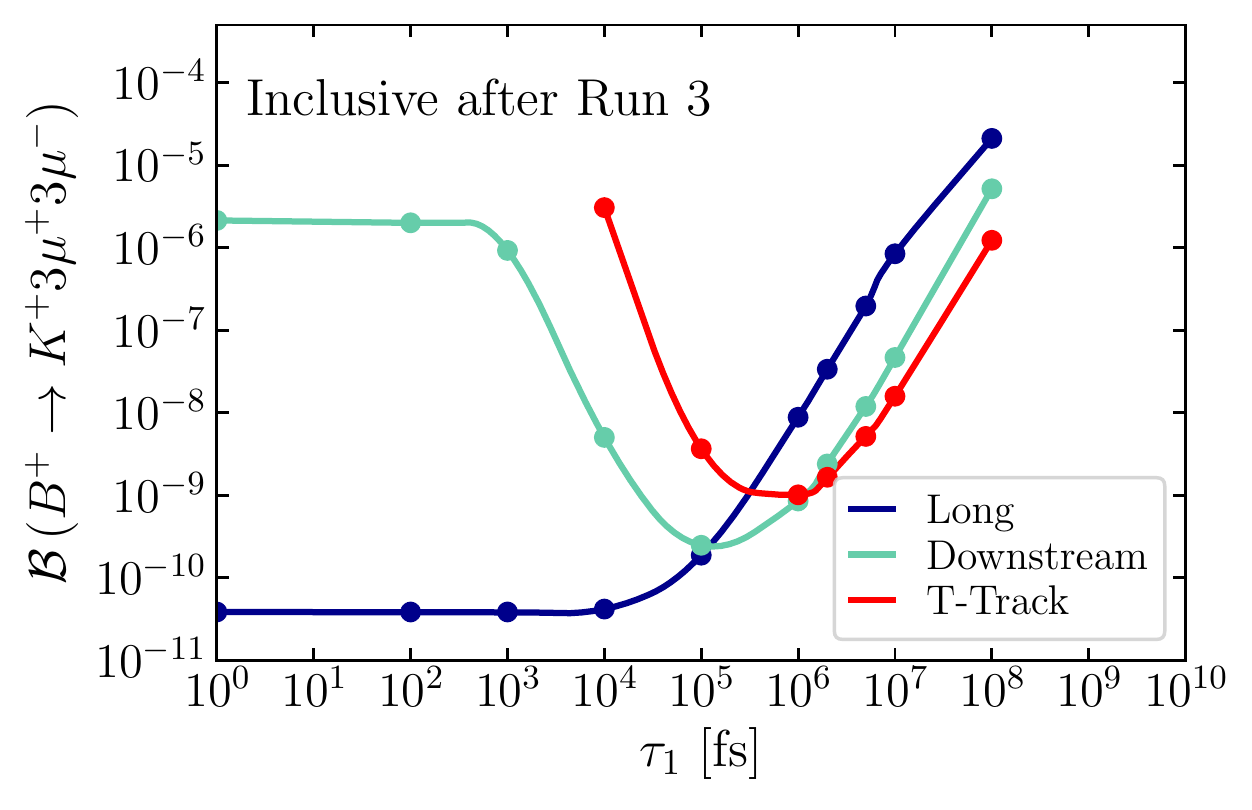}
        \includegraphics[scale=0.6]{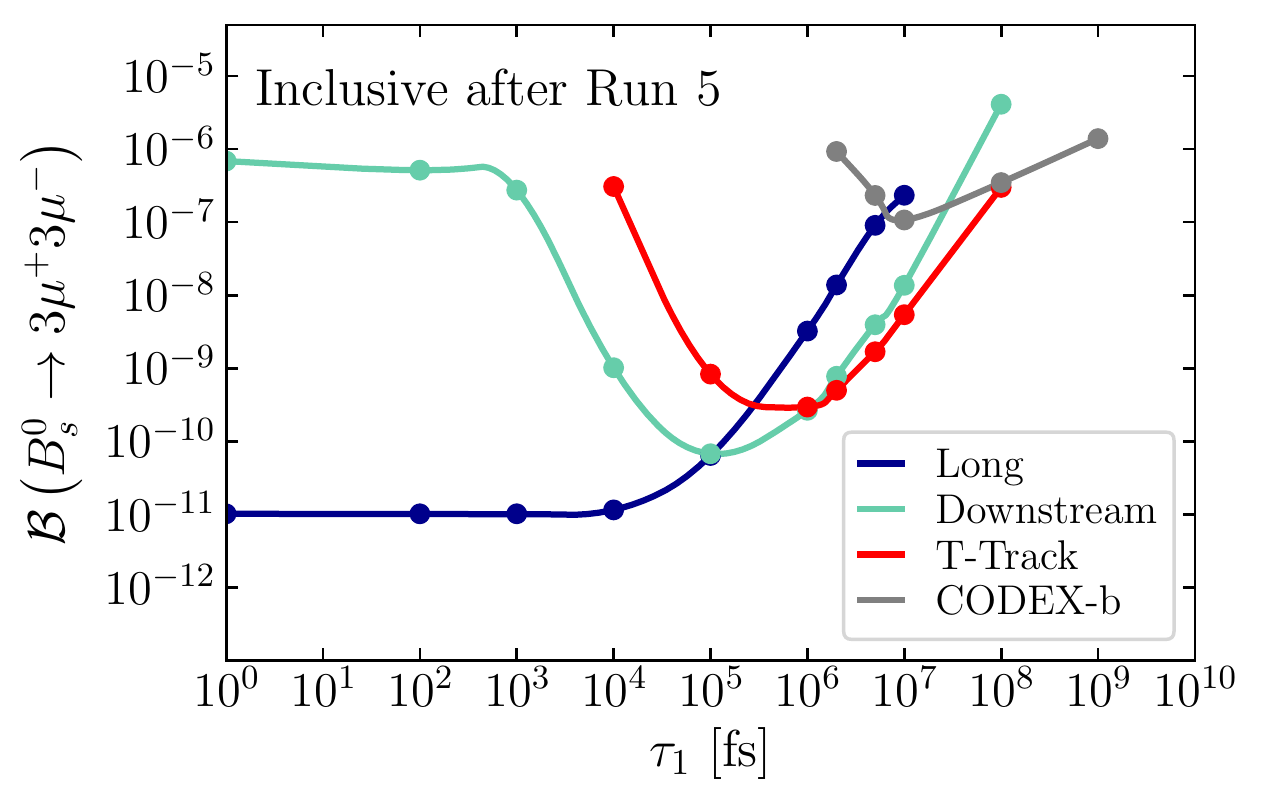}
    \includegraphics[scale=0.6]{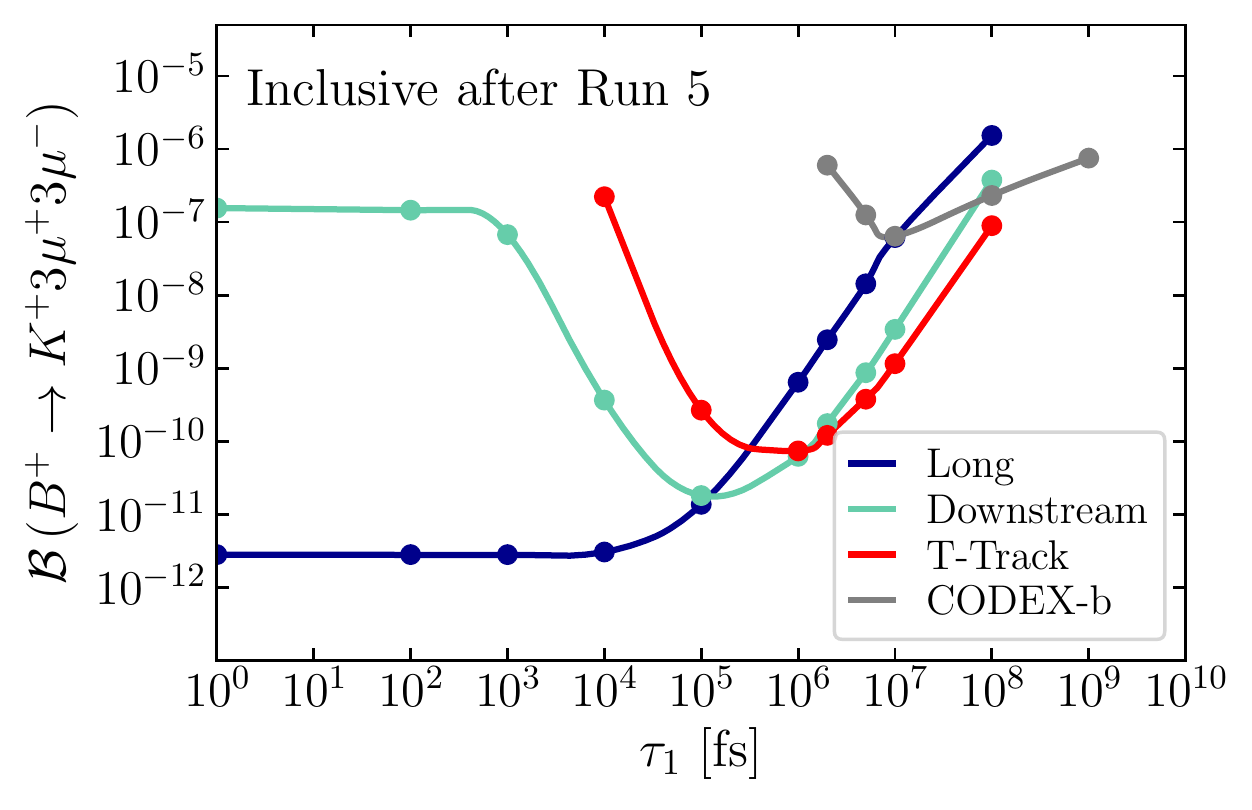}
    \caption{\it Same as in Figure~\ref{fig:lim_6mu}, but requiring only four muons in the final state. Also included are the projected bounds for the CODEX-b experiment, requiring the presence of only one muon pair inside the acceptance.}
    \label{fig:lim_4mu}
\end{figure}

An analysis of the different decay signatures is performed with simulation generated by \textsc{Pythia} 8.305~\cite{Bierlich:2022pfr}. Signal $B$ mesons are generated from $b\bar{b}$-events and decayed in many different $(m_1, m_2, \tau_1)$ configurations using a flat phase space model. Adding the momentum dependence in the meson form factors leads to negligible effects on the sensitivity (differences of around 1\% were obtained in the benchmark point used in Figs.~\ref{fig:lim_6mu} and \ref{fig:lim_4mu}).
Background processes from $pp$-collisions involving $b\bar{b}$-events and non-diffractive QCD processes are studied with \textsc{Pythia} simulation.
No attempt is made to simulate detector resolution effects. Instead these are accounted for through normalising to the published search. Similar to the published search~\cite{LHCb:2021iwr} the expected background is low, as will be discussed in Sec.~\ref{sec:bkgs} and therefore also the mass resolution is considered to make a negligible impact.

To investigate the best possible reach, an inclusive and an exclusive reconstruction approach are compared. The exclusive approach requires a reconstruction and selection of all final states and vertices of the exclusive decays, while the inclusive approach requires only the reconstruction of four muons and at least two dimuon vertices inside the detector volume. 
When considering the CODEX-b experiment, an exclusive reconstruction is found infeasible because of the large transverse displacement of $>26\,\text{m}$ and the inclusive reconstruction is relaxed by requiring at least one dimuon vertex to be inside the CODEX-b detector volume.
In the analysis of recorded LHCb data, the discriminating feature would be a clear peak at the $B$ mass in case of the exclusive reconstruction, making use of the fact that all $a_1$ candidates in the selection have a similar mass~\cite{Blance:2019ixw}. In the inclusive reconstruction the discriminating feature would be a bump in the reconstructed $4\mu$ mass where the flight direction of both dimuon candidates point to the same vertex. A clear signature in the CODEX-b detector would already be given by the existence of a vertex inside the volume.

\begin{table}[t]
    \centering
    \resizebox{16cm}{!}{
    \begin{tabular}{c|ccccc}
    \toprule
       \text{Track Type} & $x$\,[m] & $y$\,[m] & $\rho$\,[m] & $z$\,[m] & \text{Penalty Tracking Factor}\\ 
        \midrule
        Long & --- & --- & $[0,0.03]$ & $[0, 0.5]$ & $0.98$ \\ 
        
        Downstream & $[-0.75, 0.75]$ & $[-0.65, 0.65]$ & --- & $[0.6, 2.3]$ & $0.89$ \\ 
         
        T-Track &  $[-3.15, 3.15]$ & $[-2.35, 2.35]$ & --- & $[2.3, 7.6]$ & $0.70$ \\ 
         
        CODEX-b & $[26, 36]$ & $[-7, 3]$  & --- & $[5, 15]$ & $1.00$ \\ 
        \bottomrule
    \end{tabular}}
    \caption{Different track types at the LHCb and CODEX-b experiments that are studied in this work and the corresponding geometrical requirements~\cite{Bediaga:2013yyz,LHCb:2014gmu,Aielli:2019ivi}.
    The track reconstruction efficiencies are estimated according to Refs.~\cite{LHCb:2014nio,Garcia:2019jol,Ttrackreco}. For CODEX-b no penalty for the track reconstruction can be estimated yet.}
    \label{tab:tracktypes}
\end{table}

Except in the CODEX-b scenario, each considered track is required to fall inside the LHCb acceptance with an angle with respect to the beam axis between $0.01$ and $0.40\,\rm{rad}$ and has at least a transverse momentum of $250\,\rm{MeV}$ to ensure its reconstructability. For tracks inside the CODEX-b detector only the respective detector acceptance requirement is applied. 
The muon tracks are required to have a total momentum larger than $2.5\,\rm{GeV}$ to allow good muon identification.
Different decay topologies are distinguished by requiring that the $a_1\to\mu^+\mu^-$ vertices in the decay fall inside the VELO (Long), before the UT (Downstream) or SciFi (T-Track) (or in the case of CODEX-b one of the dimuon vertices falls into the CODEX-b acceptance).
On top of these requirements we include the penalty track factors presented in Table~\ref{tab:tracktypes} to model the imperfect track reconstruction.
The T-Track penalty factor, albeit lower than the others, is considered optimistic due to the very limited information from only the tracking and muon stations. These factors do not contain efficiency losses due to material interaction, which are expected to be less than $\approx10\,\%$ per track and should cancel out with respect to the reference channel in the inclusive searches. As the CODEX-b experiment is still in planning, no penalty factor is associated to the reconstruction of tracks there. However, the low occupancy of the detector would suggest a near-perfect reconstruction efficiency, thus rendering tracking an irrelevant limitation for the analysis~\cite{Aielli:2019ivi}.

\begin{sloppypar}
We reproduce the selection of the published search for the reference mode \mbox{$B^0_s\to a_1 a_1\to2\mu^+2\mu^-$} with long tracks apart from vertex reconstruction and vertex displacement requirements. In particular this implies that the requirements of the hardware trigger for muons are also applied, which was implemented during 
Run 1 and Run 2 of the LHCb experiment 
but will be removed for future data-taking periods. 
Correspondingly, we require at least one muon with $p_T>1.76\,\text{GeV}$ or at least one combination of two muons with $p_{T,1} \times p_{T,2}>(1.6\,\text{GeV})^2$~\cite{LHCb:2014set,LHCb:2018zdd}.
This results in a total efficiency of $\varepsilon_{\text{ref}}=8.4\,\%$. Removing the hardware trigger for this decay mode alone allows to increase the selection efficiency by about $20\,\%$. The total efficiency is approximately $10$ times larger than the efficiencies found in the published search. The discrepancy can be explained by additional vertexing and vertex displacement requirements which cannot be simulated with the chosen simplistic approach.
However, these requirements are likely unavoidable in any data analysis of the studied decays and therefore this factor is considered irreducible.

Under the assumption that a similar low background can be achieved with similar vertexing efficiencies as in the reference search, the expected upper limit for each scenario can be obtained as
\begin{align}
    \mathcal{B}_{\text{sig}}=\mathcal{B}_{\text{ref}}\times\frac{\varepsilon_{\text{ref}}}{\varepsilon_{\text{sig}}}\times\frac{N^{b\bar b}_{\text{ref}}}{N^{b\bar b}_{\text{sig}}}\frac{f_{s}}{f_{s,u}},
    \label{eq:BR}
\end{align}
where $\mathcal{B}_{\text{ref}}$ is the expected upper limit of the published analysis, $\varepsilon_{\text{ref}}$ and $\varepsilon_{\text{sig}}$ are the efficiencies of the signal and reference modes as determined in this work and $N^{b\bar b}_{\text{ref,sig}}$ are the effective $B$-hadron yields, obtained by multiplying the integrated luminosity of each data set with the corresponding $\sigma(pp\to b\bar{b}X)$ cross section~\cite{LHCb:2011zfl,LHCb:2013itw,LHCb:2015foc}. The hadronization fractions of the $B^0_s$ ($\sim0.1$) and the $B^+$ mesons ($\sim0.4$)~\cite{HFLAV} are labeled as $f_{s,u}$.
The integrated luminosities of the LHCb data sets are taken as $1\,\text{fb}^{-1}$ at $7\,\text{TeV}$ for the 2011 part of Run 1; $2\,\text{fb}^{-1}$ at $8\,\text{TeV}$ for the 2012 part of Run 1; and $6\,\text{fb}^{-1}$ at $13\,\text{TeV}$ for Run 2. After Run 3, Run 4 and Run 5, respectively, $23\,\text{fb}^{-1}$, $50\,\text{fb}^{-1}$ and $250\,\text{fb}^{-1}$ are expected to be collected at a center-of-mass energy of $14\,\text{TeV}$.
To obtain the $b$-quark production cross section at $\sqrt{s}=14\,\text{TeV}$, the cross section measured at $13\,\rm{TeV}$ is rescaled by a factor of $14/13$.
Thus, we assume $N^{b\bar b}_{\text{ref}}=3.9\times10^{12}$.
The linear scaling of the expected upper limit with luminosity in equation~\ref{eq:BR} is only valid under the assumption of very low background, which is discussed in detail in the following subsection.
\end{sloppypar}

The resulting expected upper limits in branching fraction for the individual decay modes are shown in Figures~\ref{fig:lim_6mu} and \ref{fig:lim_4mu}, for different assumptions of the lightest scalar lifetime. The sensitivities are presented for the different classes of tracks studied in this work. 
For the exclusive searches, by the end of Run 5 very strong sensitivities down to the level of $\mathcal{O}(10^{-11}-10^{-10})$ can be reached over a lifetime range up to $100\,\text{ps}$ if all particles are reconstructed as long tracks. Even by the end of Run 3, sensitivities of ${\mathcal{O}(10^{-10}-10^{-9})}$ can already be achieved. If all muons are reconstructed as downstream tracks, sensitivities down to $\mathcal{O}(10^{-10}-10^{-9})$ can be reached in a lifetime range between $10\,\text{ps}$ and $1\,\text{ns}$. Considering all muons reconstructed as T-Tracks allows to reach a moderate increase in efficiency (a factor of $\approx 3$) and thus improvement in the expected sensitivity for lifetimes between $1$ and $100\,\text{ns}$ for the final states discussed in this work. For these searches, extra sensitivities might be gained in overlap regions when mixed track types are considered: $e.g.$ in the case of $B_s^0\to 3\mu^+ 3\mu^-$, if two dimuon pairs are reconstructed as downstream tracks and the other two muons as T-Tracks.

While not being selective to the final state and thus not as fit to distinguish between models, only reconstructing four muons instead of the full exclusive final states can improve the search sensitivity by about an order of magnitude.
Reconstructing dimuon vertices in the CODEX-b experiment could improve the sensitivity to
decays at very large lifetimes between $100\,\text{ns}$ and $1\,\mu\text{s}$, allowing to test branching fractions
of $\mathcal{O}(10^{-7}-10^{-6})$ by the end of Run 5.
Therefore, a good complementary could be achieved between the LHCb and Codex-b experiments in the large lifetime regime.
It should be noted however that if CMS is able to maintain a dedicated dimuon trigger as the one used in Ref.~\cite{CMS:2021sch}, the limits from displaced vertex analyses could surpass the CODEX-b reach in the very near future; see Section~\ref{sec:L}. In this regard, a detailed study comparing the sensitivity of the several LHC experiments to the hierarchical sectors would be very interesting.

In Appendix~\ref{sec:maps}, the mass-dependent expected upper limit distributions are included as a function of the lifetime. A good sensitivity can be achieved across the whole phase space with variations of less than a factor of 2. Thus the projections made in Figures~\ref{fig:lim_6mu} and \ref{fig:lim_4mu} are valid with this uncertainty despite being made at a single benchmark mass.
A clear structure can be observed.
Larger efficiencies (and correspondingly smaller expected limits) are observed for high and low $m_1$ values. This structure is caused by the requirements on the transverse momentum of the muons. Products of heavy particle decays have a larger spread of transverse momenta than those of light particles. In the limit of low $m_1$, each pair of muons is produced collinearly and their transverse momentum is defined by the heavy ancestor, ultimately the $B$-meson. For high $m_1$, the $a_1$ scalars themselves are heavy and thus a source of larger muon $p_T$. At high $a_1$ lifetimes, the efficiency increases continuously with increasing $m_1$ as the geometrical requirement of the detector components become more important. The higher $m_1$, the less its boost away from the direction of the $B$; thus the fraction of scalars (and consequently muons) that can be reconstructed inside the LHCb acceptance increases.

\subsection{Expected backgrounds}
\label{sec:bkgs}
For the proposed searches with the LHCb experiment several sources of background have to be considered.
A dominant contribution typically arises from random track combinations.
As demonstrated in the recently published LHCb search for four muon final states, it is possible to effectively remove these backgrounds with vertexing and muon identification requirements.
If that is true for the four muon final state, the same will hold for the decays with even more muons in the final state.
Additionally, the behaviour of possible (dominantly combinatorial) background processes has been studied with inclusive $pp\to b\bar b\text{X}$ and non-diffractive QCD simulation, that was reconstructed with the decay chain $B^0_s\to a_1a_2\to2\mu^+2\mu^-$. The same requirements (including muon identification, taking into account the misidentification efficiencies presented in Refs.~\cite{LHCb:2013lvx,LHCb:2014set}) as discussed in Section~\ref{ref:sec-limits} are applied, imposing additionally the $B$ candidate mass to lie between $4.5$ and $6\,\text{GeV}$ and a distance of closest approach (DOCA) between the scalars and each dimuon pair forming a vertex to be $<1\,\text{mm}$.
It is found that $>99\,\%$ of the backgrounds (mainly random track combinations) come from processes inside the VELO. Only $<1\,\%$ of the background processes have at least one dimuon candidate decaying downstream of the VELO. Thus if the background processes inside the VELO can be controlled, the ones outside the VELO can be suppressed even better.

\begin{figure}
    \centering
    \includegraphics[width=0.32\textwidth]{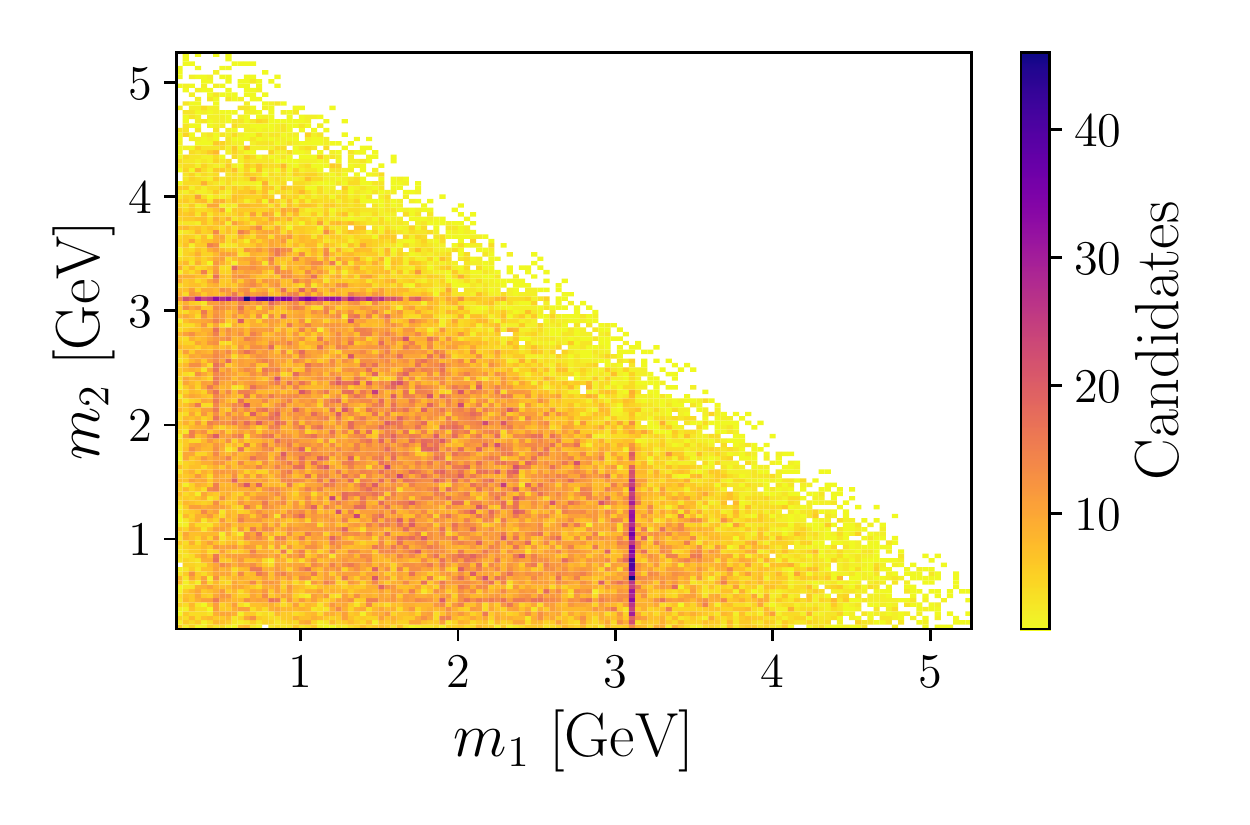}
    \includegraphics[width=0.32\textwidth]{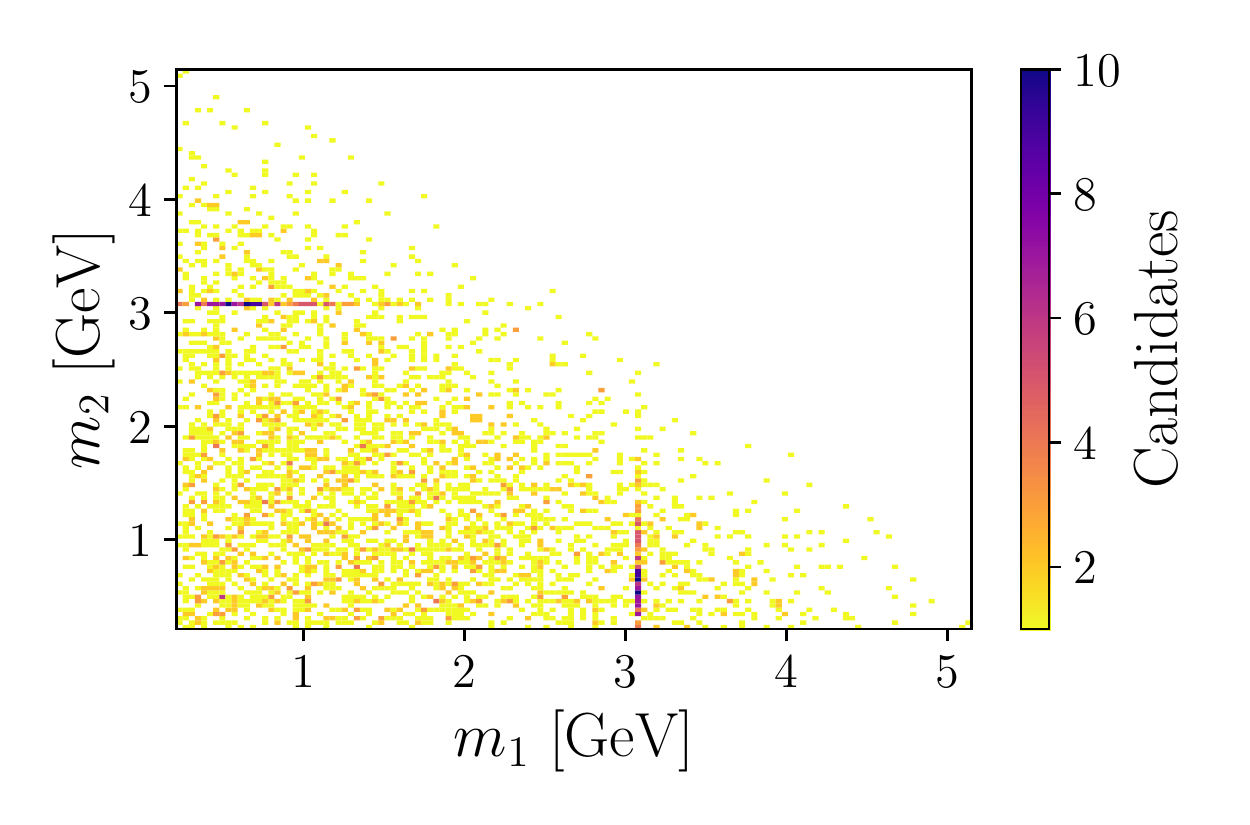}
    \includegraphics[width=0.32\textwidth]{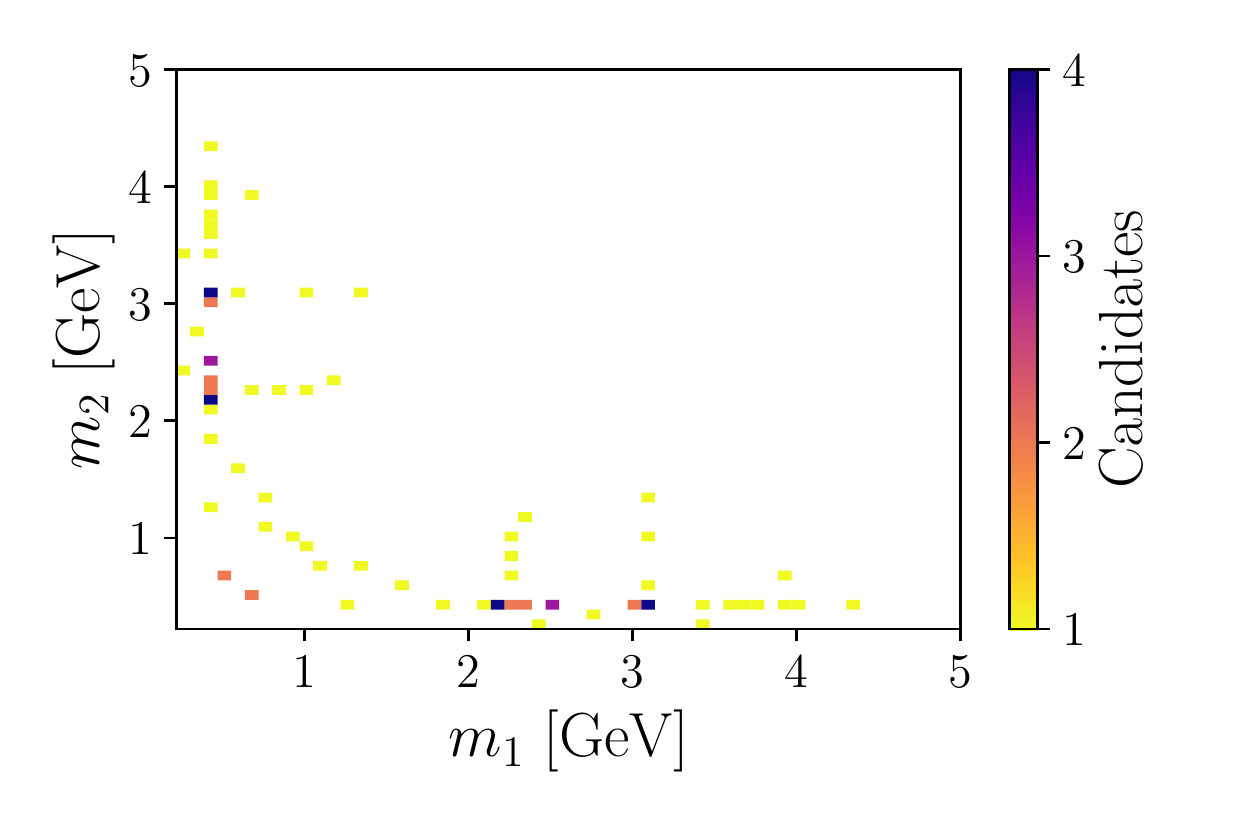}
    \caption{\it Dimuon mass distributions in simulated $pp\to b\bar b\text{X}$ events reconstructed as $B^0_s\to a_1a_2\to2\mu^+ 2\mu^-$. In the left panel, only the requirements discussed in Section~\ref{ref:sec-limits} are applied. In the central panel, strong vertex requirements are applied. In the right panel, it is required instead that one of the dimuon systems has a significant flight distance. The vertexing strongly reduces the $K^0_{\text{S}}$ resonance at $\approx500\,\text{MeV}$, while the flight distance requirement removes the $J/\psi$ resonance around $\approx3\,\text{GeV}$.}
    \label{fig:selection}
\end{figure}

However, there can be particle decays with intermediate dimuon or dipion resonances that might mimic the signal processes. 
These are dominantly real dimuon resonances like $J/\psi(1S)$, $\Psi(2S)$ and $\phi(1020)$, as well as $K^0_\text{S}\to\pi^+\pi^-$ decays where the pions are misidentified as muons. Although the latter are strongly suppressed by the muon identification requirements, they are so abundant that they can still fake a dimuon signal.
The relevant processes can either come directly from a $B$-meson, such as $B^0_s\to J/\psi\phi$ and $B^0_s\to J/ \psi K^0_{\text{S}}$, or be randomly combined.
Explicit $B$ decays like the above mentioned will thus have to be vetoed; however they constitute only a very small fraction of the phase space. For random combinations of the dimuon resonant processes, again making strict vertexing requirements and imposing a significant lifetime for one of the dimuons will allow to successfully erase those backgrounds in a similar way as for backgrounds from random muon combinations, as demonstrated in Figure~\ref{fig:selection}. Despite the muon identification requirements, clear contributions from $J/\psi(1S)$ and $K^0_\text{S}$ resonances persist. The former can be removed by requiring a significant distance (in this case $50\,\text{mm}$) between the reconstructed $B$ vertex and one scalar vertex. The latter can be removed requiring a good vertex between the two scalars and each dimuon candidate, simulated by $\text{DOCA}<0.1\,\text{mm}$. This is effective, since the $K^0_\text{S}$ particles do not tend to be produced in the same vertex.
Additional backgrounds might come from $J/\psi\to2\mu^+2\mu^-$ decays~\cite{Chen:2020bju}, which can be vetoed efficiently for the search for the low-lifetime configurations and will be negligible as soon as a significant lifetime for at least one of the dimuon candidates is required.

The exclusive study of the six muon final states will not suffer from the previous limitations allowing to avoid any veto, as there is no SM process that can be misidentified as the signal with rates close to the reachable sensitivities ($e.g.$ decays like $B^0_s\to J/\psi\phi\phi$ have branching fractions of the order of $10^{-13}$~\cite{pdg}).

Finally, important backgrounds mimicking long-lived signals can come from charged-particle interactions with the detector material. No attempt is made to simulate these backgrounds. However, the detector geometry is well known and data-driven tools~\cite{Alexander:2018png} have been developed to efficiently suppress material interaction processes for data already collected with the LHCb detector, which will be transferable to future detector modifications.

 CODEX-b is considered a background-free experiment. The strong background suppression is achieved by a significant displacement and shielding with respect to the $pp$-collision point. Furthermore, an active veto against neutral long-lived-particle production in the shielding material is employed. This results in the fact that any dimuon vertex registered in the CODEX-b volume can be considered as signal.

Altogether, these arguments show that a search with negligible background can be performed, validating the use of equation~\ref{eq:BR}.

\section{Model interpretation}~\label{sec:model}
Having estimated the LHCb sensitivity to the rare processes under study, the limits on the parameter space of the model presented in equation~\ref{eq:Lsetup} can be readily obtained. To this end, the stringent upper limits obtained in the previous section (associated to the inclusive search) are used.

\begin{figure}[t]
    \centering
    \includegraphics[scale=0.62]{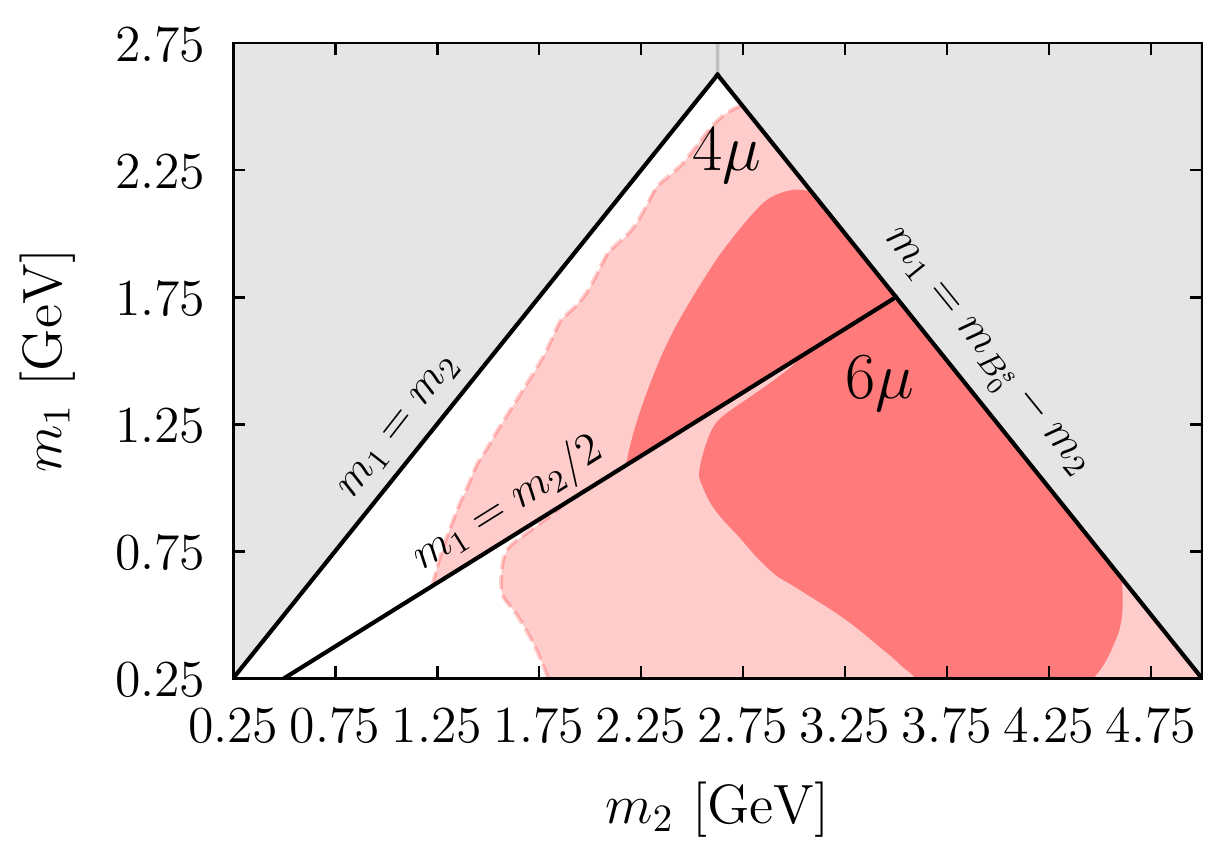}
    \includegraphics[scale=0.62]{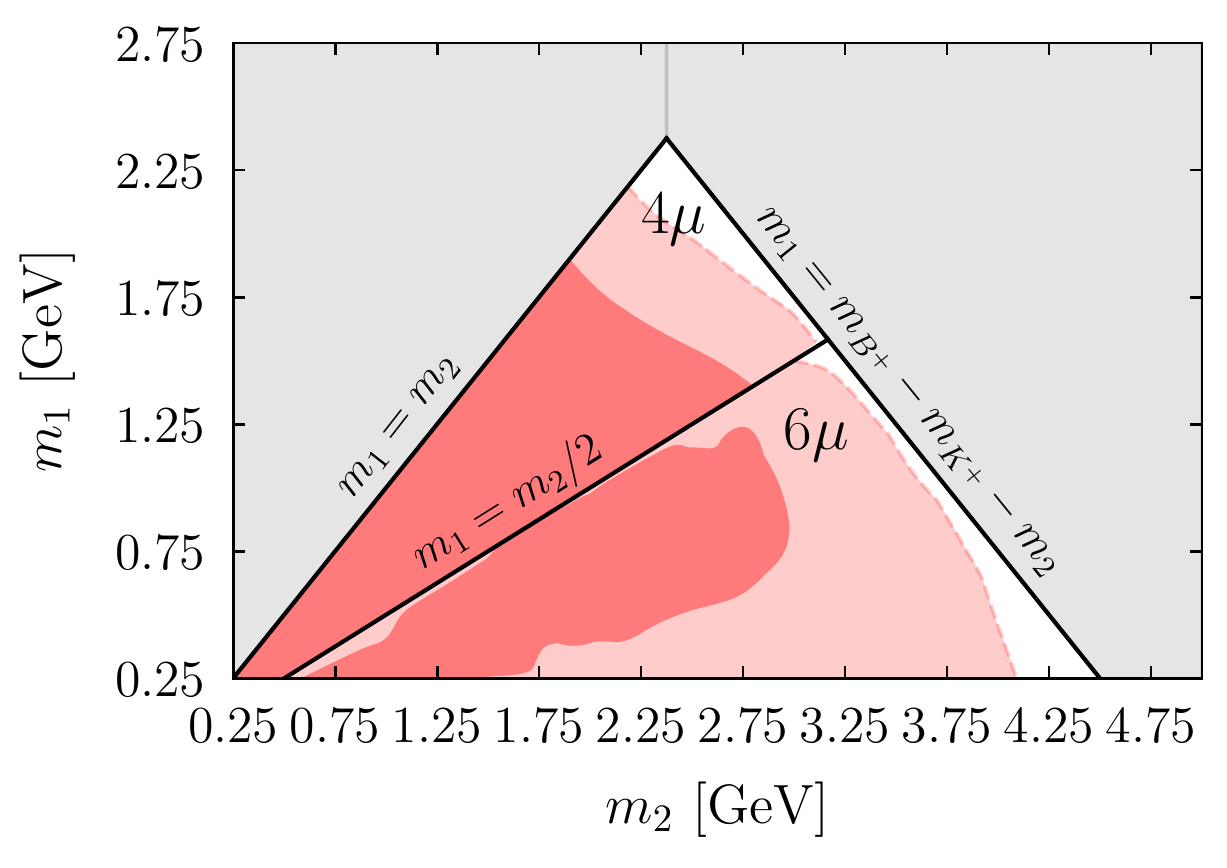}
    \caption{\it (Left) Light scalar masses that could be probed with the analysis proposed in this work in the channels $B_s^ 0\to 2\mu^+ 2\mu^-$ ($m_2 < 2 m_1$ region) and $B_s^ 0\to 3\mu^+ 3\mu^-$ (${m_2 > 2 m_1}$ region).
The dark and light red lines enclose the parameter space that could be probed, respectively, at Run 3 and Run 5 of LHCb. 
    The lighest particle lifetime is fixed to ${\tau_1 = 100\,\text{ps}}$, $g_{12} = 1.5$ and the chosen resonance parameters are signaled by a star in Figure~\ref{fig:MV}. Only the upper limits derived in section~\ref{ref:sec-limits} from downstream tracks were considered. (Right) The same analysis applied to ${B^+ \to K^+ 2\mu^+2\mu^-}$ and ${B^+ \to K^+ 3\mu^+3\mu^-}$.
    }
    \label{fig:M1M2}
\end{figure}

\begin{figure}[t]
    \centering
    \includegraphics[scale=0.62]{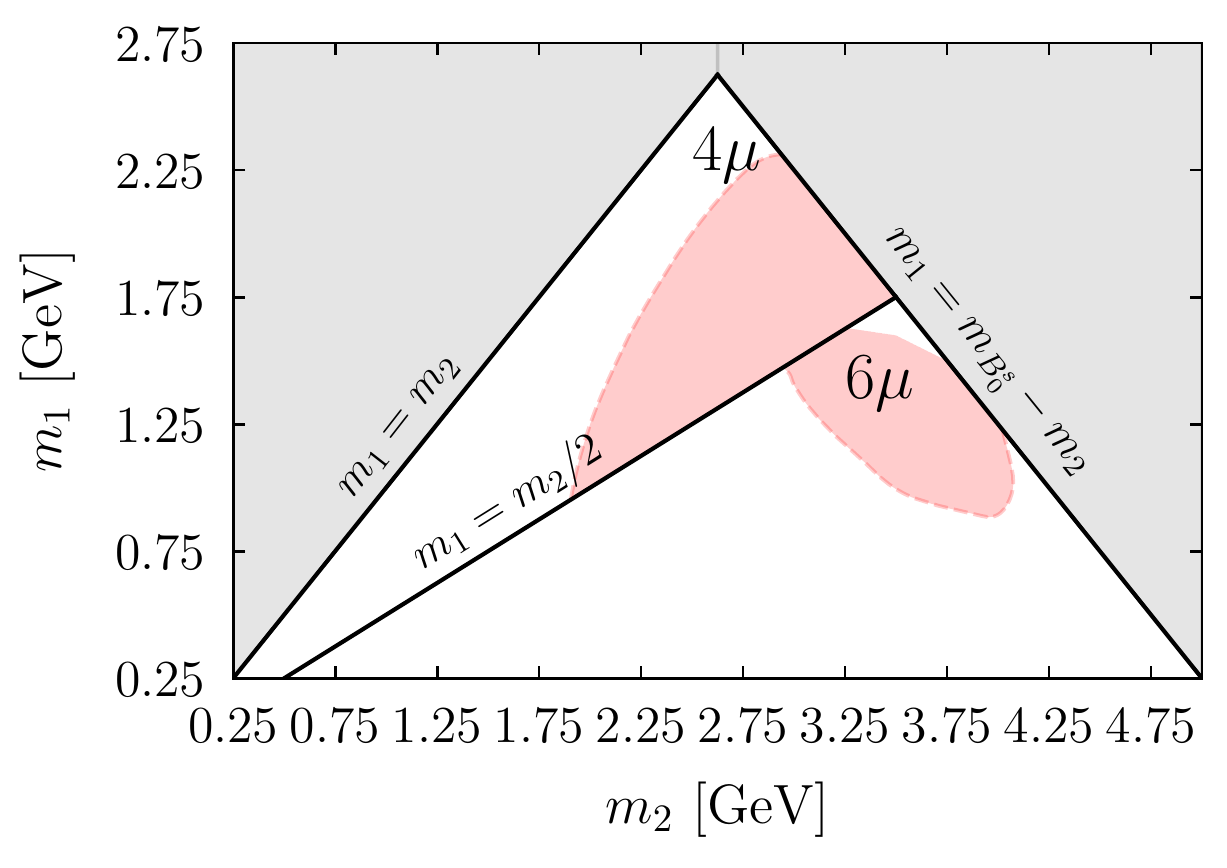}
    \includegraphics[scale=0.62]{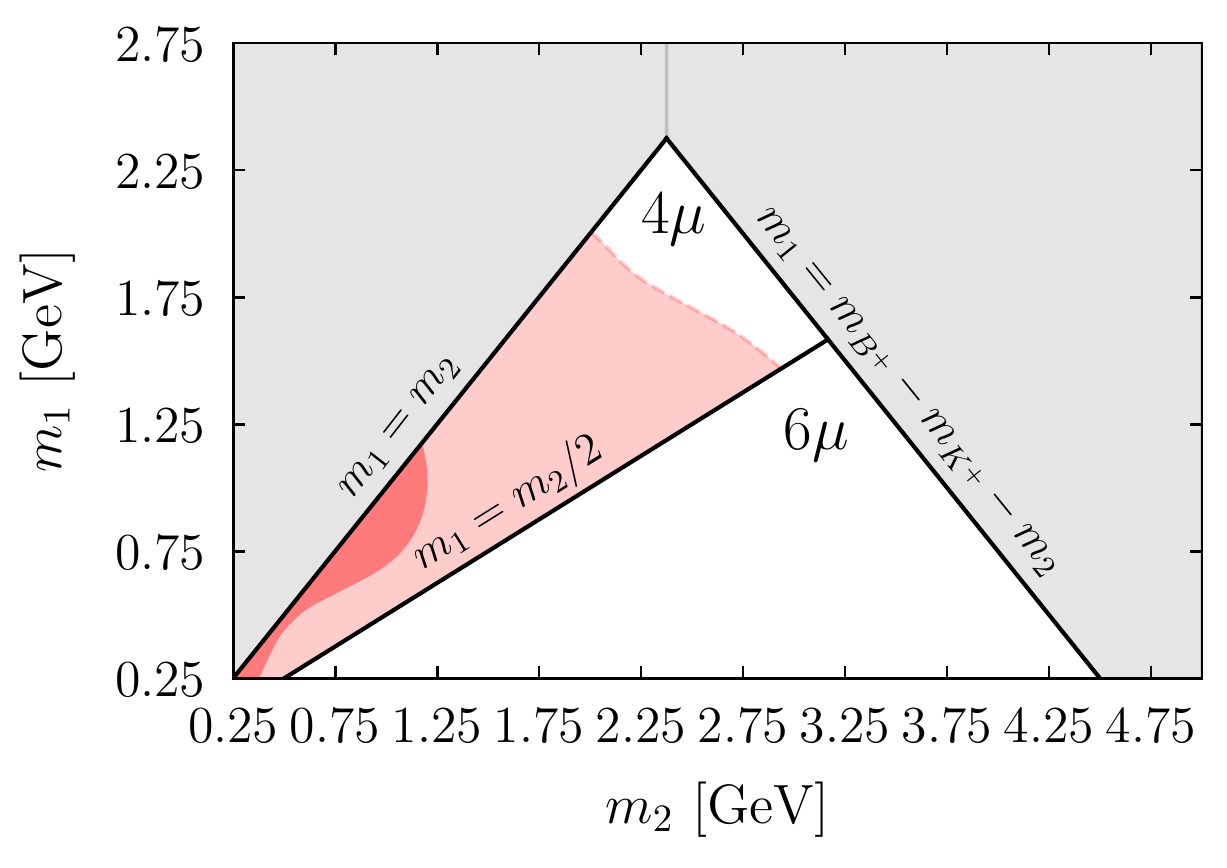}
    \caption{\it Same as Figure~\ref{fig:M1M2} but considering $\tau_1 = 10$\,ns. This time only the upper limits derived in Section~\ref{ref:sec-limits} from T-tracks are used.
    }
    \label{fig:M1M2_TT}
\end{figure}

In Figure~\ref{fig:M1M2}, the reach of the proposed analysis to the hierarchical light sector, comprising both the short- and long-lived singlets, is represented in the $(m_1,m_2)$ plane. The lightest scalar lifetime is fixed to $\tau_1 = 100\,\text{ps}$ in this example.
As can be seen, the central area of the phase space could be tested in either of the two channels, $B_s^0$ or $B^+$, to multiple muons.  On the other hand, these have a complementary role in probing the borders of the phase space. Due to the higher production rate of $B^+$ mesons, the analysis applied to the channel ${B^+\to K^+ 3\mu^+ 3\mu^-}$ could lead to the strongest limits, being able to probe the parameter space where $m_1\sim m_2$ that is left unconstrained by the two-body decay of the $B$. Along $m_2$, the efficiencies are essentially constant allowing to understand straightforwardly the behaviour along the horizontal lines of the plots. Since $\Gamma (B_s^0\to a_1 a_2)$ grows with the scalar masses~\cite{Blance:2019ixw}, the potential exclusion region is found for large $m_2$. On the other hand, the three-body decay $B^+ \to K^+ a_1 a_2$ is kinematically suppressed for large $m_{1,2}$ which explains the opposite behaviour of the panels in Figure~\ref{fig:M1M2}.

The constraints in the figure above were obtained assuming the reconstruction of downstream tracks. Had long tracks been used instead for the assumed lifetime of $a_1$, 
the limits 
at Run 5 in the kinematic region $m_2 > 2 m_1$ would be comparable to the Run 3 exclusion area in the left panel of Figure~\ref{fig:M1M2}.
 Moreover, the corresponding region in the right panel would remain completely unprobed.
This clearly shows the benefit of introducing new trigger strategies with alternative tracking approaches for probing the non-minimal scalar sectors. 

For even larger scalar lifetimes, analyses making use of T-Tracks could possibly lead to the strongest limits on the model parameter space, as represented in Figure~\ref{fig:M1M2_TT}. Assuming the same model parameters as in that figure, a downstream track based analysis would be insensitive to the entire kinematic region where $m_2 > 2 m_1 $.

Another remark is in order regarding the potential solution to the $(g-2)_\mu$ anomaly in this setup.
The parameter space compatible with such solution in the next-to-minimal CHM (with only one pNGB) was recently probed in the recent analysis by LHCb focused on final states with four prompt muons~\cite{LHCb:2021iwr}. 
However, if a light $a_2$ explains this anomaly while  decaying mostly into a long-lived scalar, such parameter space could only be probed at future LHC Runs using the reconstruction of tracks only in subsystems of the LHCb detector.

\begin{figure}[t!]
    \centering
    \includegraphics[scale=0.54]{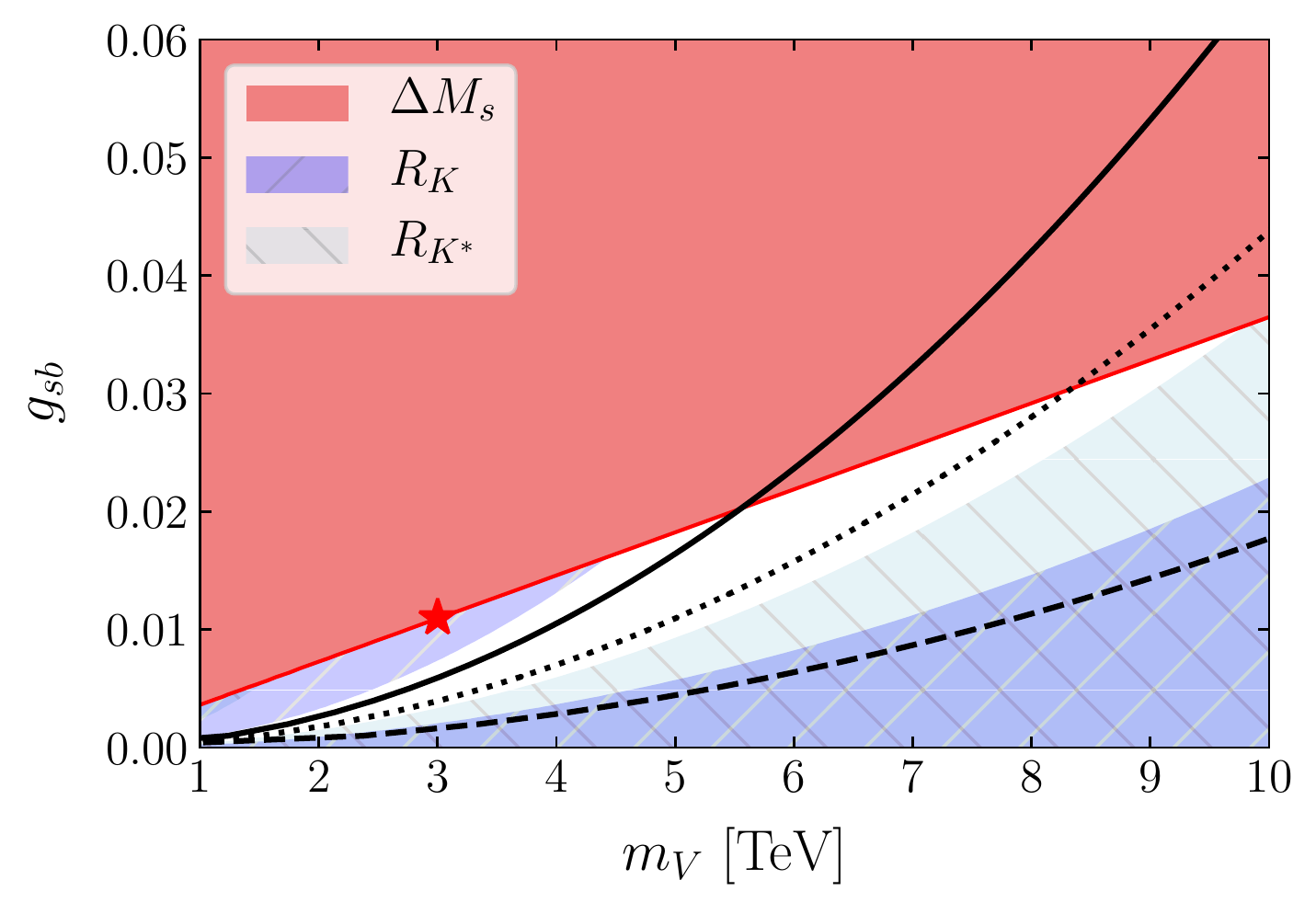}
    \includegraphics[scale=0.54]{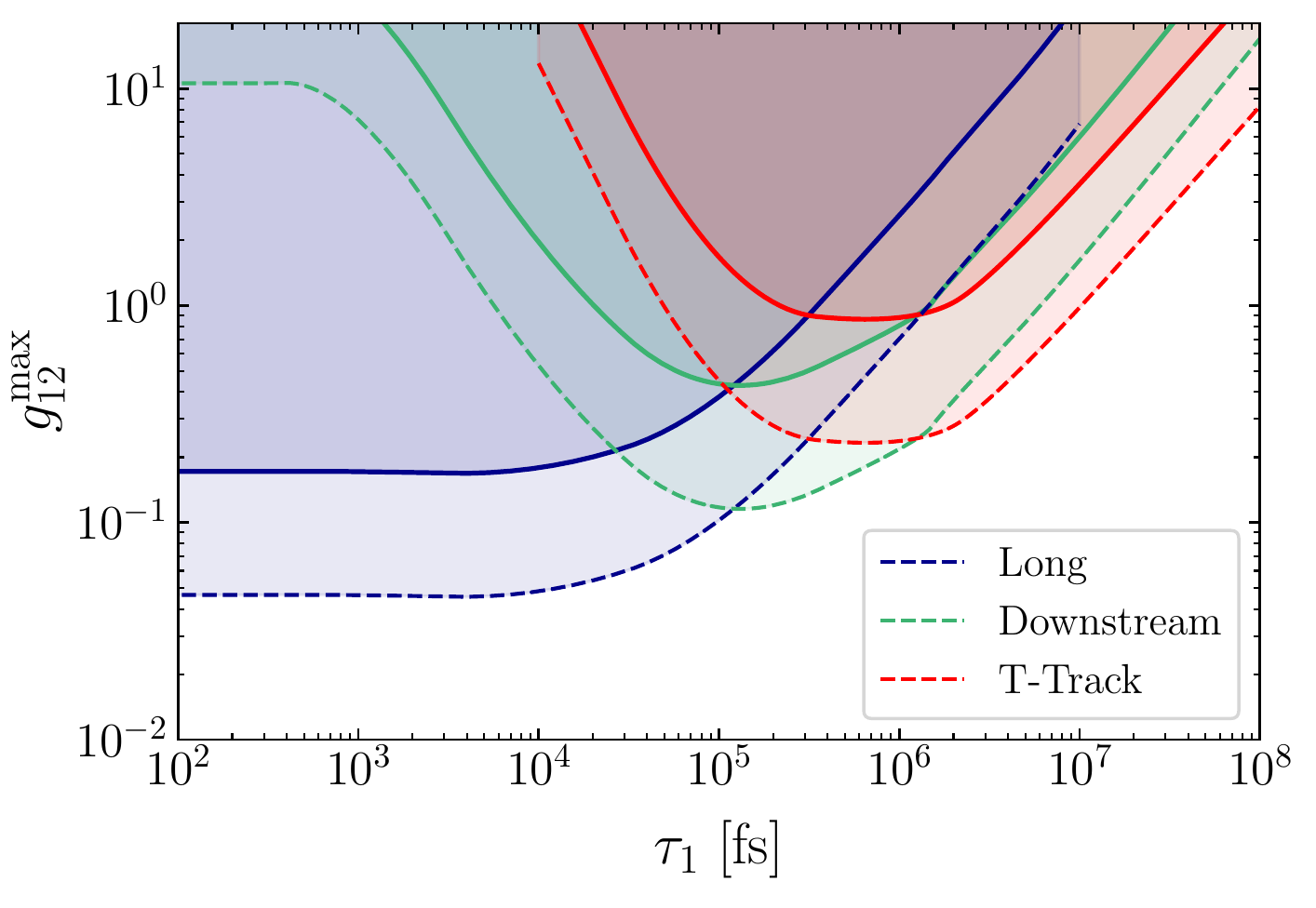}
    \caption{\it (Left) The minimum vector-quarks coupling that could be tested with the search proposed in this work in the channel $B^+\to K^+ 3\mu^+3\mu^-$. The light masses, couplings and lifetimes are assumed to be $(m_1,m_2)=(1.0,2.5)$  GeV, ${g_{12} = 1.5}$ and ${\tau_1 = 1}$ ns. The solid, dotted and dashed lines show, respectively, the projected limits after Run 3, Run 4 and Run 5 of the LHCb experiment. (Right) The minimum vector-scalar coupling that could be tested with the same analysis, as a function of the lightest scalar lifetime.
    The same notation and benchmark masses as in the left panel are considered, as well as the mass and vector coupling marked with a star therein.}
    \label{fig:MV}
\end{figure}

Next, the constraints on the heavy resonance mediating the $B$ decays into the light scalars are discussed, assuming the presence of flavour violation in the quark sector.
Strong bounds from meson mixing constrain the plane $(m_V,\,g_{sb})$, which are represented (in red) in the left panel of Figure~\ref{fig:MV}. Such exclusion region was found using the recent weighted averages reported in Ref.~\cite{DiLuzio:2019jyq} for the SM prediction and the $2\sigma$ limits on $\Delta M_s$ presented in Ref.~\cite{LHCb:2021moh}. Superimposed (in blue) are the $2\sigma$ regions where a $V$ explanation to the $R_{K^{(*)}}$ anomalies is ruled out. The tree level contribution of $V$ to these observables was computed with 
 EOS~\cite{vanDyk:2021sup} and the updated results from LHCb~\cite{LHCb:2021trn,LHCb:2017avl} were considered to constrain this contribution.
 
By performing the analysis proposed in this work, a significant region of the phase space where a combined solution to the neutral anomalies is possible could be tested already at Run 4, even if the vector decays mostly into the hidden long-lived sector.
In turn, at Run 5, also the $V$ explanation to $R_K$ alone could be entirely probed. (Such conclusion would hold if the four-muon search was applied to the $B_s\to a_1 a_2$ channel, although the limits are typically weaker by a factor of $\sim 3-4$ due to the lower hadronisation fraction of the $B^0_s$ mesons.)

Assuming the maximum allowed $g_{sb}$ coupling for a TeV vector resonance, constraints can be instead applied on its coupling to the light pNGBs, as shown in the right panel of Figure~\ref{fig:MV}. 
Such coupling is a smoking gun feature of the composite scenario, being predicted to be $g_{12} \sim g_* \sim \mathcal{O}(1)$. 
Furthermore, the latter can be entirely determined from the CCWZ prescription~\cite{Panico:2015jxa} and is independent of the explicit breaking details of the model.
It is evident that for lifetimes of the lightest scalar $\tau_1 \gtrsim 1\,\text{ns}$, the composite nature cannot be probed at LHCb by searching only for long tracks.
This conclusion is, in good approximation, independent of the benchmark masses; see Appendix~\ref{sec:maps}. Codex-b could provide additional sensitivity to the composite coupling for $\tau_1 > 100\,\text{ns}$; however, for the chosen benchmark masses in Figure~\ref{fig:MV}, only $g_{12} \gtrsim \mathcal{O}(10)$ could be reached.

\section{Summary}~\label{sec:summary}
In this work, the signatures of hierarchical BSM sectors at LHCb have been studied in detail, with focus on the low-energy predictions of the coupling between a heavy resonance and a Goldstone boson. In scenarios where the Goldstone sector is non-minimal and the lightest scalar is long-lived, such coupling would trigger novel $B$ decays into multiple displaced muons that evade the target of current searches.

Therefore, novel analyses were proposed considering realistic configurations of the future LHCb detector with access to lower momentum thresholds and exploring new track reconstruction techniques that allow sensitivity to large scalar lifetimes. In particular, by considering long, downstream and T-tracks in the analyses, scalar lifetimes spanning seven orders of magnitude could be probed with expected sensitivities reaching ${\mathcal{B}(B\to 3\mu^+3\mu^- (K)) \sim \mathcal{O}(10^{-9})}$ considering the full LHCb data set. The upper limits set on the relevant branching fractions are expected to become weaker by more than one order of magnitude if only long tracks are included in future analyses targeting the long-lived scalar.
To complement the discussion, the reach of the proposed CODEX-b experiment has also been studied, providing the strongest constraints for lifetimes $\gtrsim 100$\,ns. 

If observed, the $B$ decay processes under study could be a hint of a composite Higgs sector in the TeV region, where a strong resonance-Goldstone coupling is predicted. The high multiplicity final states could be in this case a consequence of a non-minimal UV symmetry and hence provide information about the pattern of spontaneous symmetry breaking of the composite sector. In the same way, the couplings and lifetimes of the Goldstone particles under study could be an indication of the quantum numbers of the lightest resonances that generate the composite-elementary mixings, as illustrated in Appendix~\ref{sec:embedding}. Interpreting the obtained limits in this context, compositeness couplings of $\mathcal{O}(1)$ could be tested up to lifetimes of $\mathcal{O}(100)$\,ns with the inclusion of incomplete tracks at the high-level LHCb trigger. The parameter space of the vector resonance motivated by the flavour anomalies could also be tested, even if it couples mostly to the long-lived scalar.

Nevertheless, the multi-muon displaced signatures could arise in other BSM scenarios, considering for example other portals to the singlet particles~\cite{Batell:2009jf}. Furthermore, such signatures could also appear in less simplified versions of the typical
co-annihilation~\cite{Baker:2015qna,Khoze:2017ixx} and freeze-in dark matter~\cite{Hall:2009bx,Guedes:2021oqx} scenarios, involving a non-minimal dark sector with sizable couplings to leptons. Other examples include
dark QCD sectors~\cite{Pierce:2017taw} and non-minimal dark photon models~\cite{Acevedo:2021wiq}.
To interpret the upper limits in other frameworks, the efficiency maps are provided across masses and for different lifetimes of the scalar particles in Appendix~\ref{sec:maps}.

\section*{Acknowledgements}
The authors are very thankful to Simon Knapen, Dean Robinson, Mikael Chala and the LHCb physics coordination for the interesting discussions.
MR has received support from the European Union’s Horizon 2020 research and
innovation programme under the Marie Skłodowska-Curie grant agreement No 860881-
HIDDeN.
TM has received support from the Spanish Research State Agency through the Juan de la Cierva Formación grant FJC2020-045496-I. The work of XCV is supported by MINECO (Spain) through the Ram\'{o}n y Cajal program RYC-2016-20073 and by XuntaGAL under the ED431F 2018/01 project. IGFAE members have received financial support from Xunta de Galicia (Centro singular de investigación de Galicia accreditation 2019-2022), by European Union ERDF, and by the ``María de Maeztu'' Units of Excellence program MDM-2016-0692 and the Spanish Research State Agency.

\appendix

\newpage
\bibliographystyle{BiblioStyle.bst}
\providecommand{\href}[2]{#2}\begingroup\raggedright\endgroup
\newpage

\section{Composite embedding}~\label{sec:embedding}

In CHMs~\cite{KAPLAN1984183,KAPLAN1984187}, the Yukawa couplings and the potential of the Goldstone particles are generated at tree level and at one-loop, respectively, from the mixing between heavy fermions of the composite sector and those of the SM~\cite{KAPLAN1991259,Panico:2015jxa}. 
Therefore, if the quarks transform in representations of the global group that preserve the shift-symmetry of the Goldstone particles, their masses are expected to be small, of order $y^\ell f/(4\pi)$ with $f$ denoting the compositeness scale. The lepton sector can be the main source of symmetry breaking in these cases.

To discuss an explicit example, the $SO(7)/SO(6)$ CHM~\cite{Chala:2016ykx,Balkin:2017aep,DaRold:2019ccj,Blance:2019ixw,Ramos:2019qqa} is considered assuming that the 
left-handed (LH) leptons are embedded in the $\mathbf{27}_L=\mathbf{1}_L \oplus \mathbf{6}_L \oplus \mathbf{20}_L$ representation of $SO(7)$
while the right-handed (RH) ones transform both in the $\mathbf{7}_R=\mathbf{1}_R\oplus \mathbf{6}_R$ and in the $\mathbf{1}_R^\prime$. (In the previous equations, the decomposition rules under the unbroken group are shown explicitly.)
The most general embedding for the LH leptons in the symmetric representation is given by
\begin{equation}
L_L = \begin{pmatrix}
0 & 0 & 0 & 0 & -\frac{\sqrt{2}\theta_1}{2+\sqrt{2}} e_L & \ii \gamma_1 e_L & \ii e_L\\
0 & 0 & 0 & 0 & \ii \frac{\sqrt{2}\theta_1}{2+\sqrt{2}} e_L & \gamma_1 e_L & e_L \\
0 & 0 & 0 & 0 & -\frac{\sqrt{2}\theta_1}{2+\sqrt{2}} \nu_L & \ii \gamma_1 \nu_L & \ii \nu_L \\
0 & 0 & 0 & 0 & -\ii \frac{\sqrt{2}\theta_1}{2+\sqrt{2}} \nu_L & -\gamma_1 \nu_L & -\nu_L \\
-\frac{\sqrt{2}\theta_1}{2+\sqrt{2}} e_L & \ii \frac{\sqrt{2}\theta_1}{2+\sqrt{2}} e_L & -\frac{\sqrt{2}\theta_1}{2+\sqrt{2}} \nu_L & -\ii \frac{\sqrt{2}\theta_1}{2+\sqrt{2}} \nu_L & 0 & 0 & 0 \\
\ii \gamma_1 e_L & \gamma_1 e_L & \ii \gamma_1 \nu_L & -\gamma_1 \nu_L & 0 & 0 & 0 \\
\ii e_L & e_L & \ii \nu_L & -\nu_L & 0 & 0 & 0
\end{pmatrix}\,;
\end{equation}
while $E_R^{\mathbf{7}} = (0,0,0,0,\ii \theta_2 e_R, -\gamma_2 e_R, e_R)$. Without loss of generality, the parameters $\gamma_{1,2}$ and $\theta_{1,2}$ are taken to be real.

The scalar potential in this model is constructed out of three $SO(6)$ invariants, the $\mathbf{1}_L \times \mathbf{1}_L,~\mathbf{6}_L\times \mathbf{6}_L$ and $\mathbf{1}_R \times \mathbf{1}_R$; the corresponding UV constants are denoted by $c_1$, $c_2$ and $c_3$. At leading order,
\begin{align}
V =&  a_2 f^3 (4 \gamma_1 c_2 - \gamma_2 c_3) - \frac{f^2}{8} \bigg[ 4 (a_1^2 + a_2^2) (4 c_2 + c_3) - 4 c_2 ( 4 \gamma_1^2 a_2^2 + \theta_1^2 a_1^2) - 4 c_3 ( \gamma_2^2 a_2^2 + \theta_2^2 a_1^2) \bigg] \nonumber\\
& + \frac{f}{2} a_2 \bigg[ 4 \gamma_1 c_1 h^2 - 4 \gamma_1 c_2 (a_1^2 + a_2^2 + 3 h^2) + \gamma_2 c_3 (a_1^2 + a_2^2 + h^2) \bigg] \nonumber \\
& -h^2 (c_1 - 2 c_2) (a_1^2 + a_2^2 - \gamma_1^2 a_2^2 - \frac{1}{4} \theta_1^2 a_1^2) + \dots
\end{align}
where the dots encode interactions involving the Higgs boson solely, which are assumed to be determined from the couplings of the SM quarks to the composite operators.

\begin{figure}[t]
    \centering
    \includegraphics[scale=0.55]{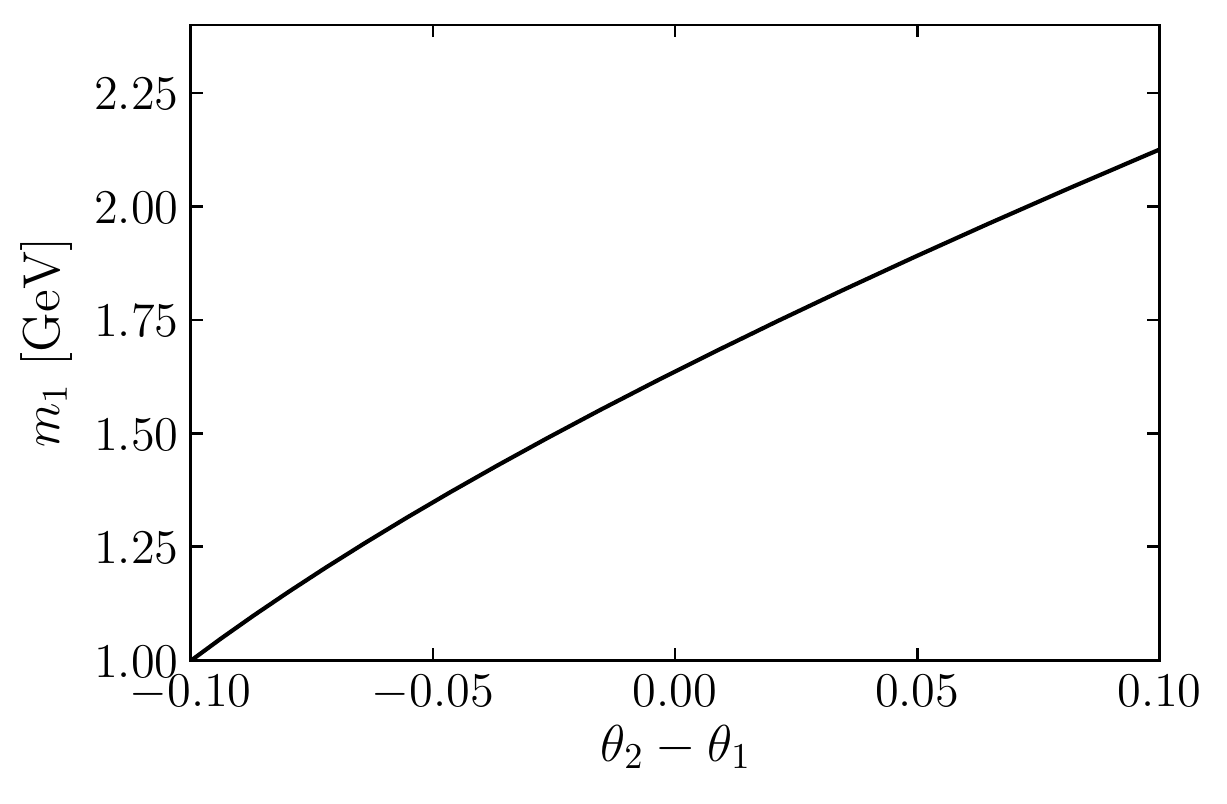}
 \includegraphics[scale=0.55]{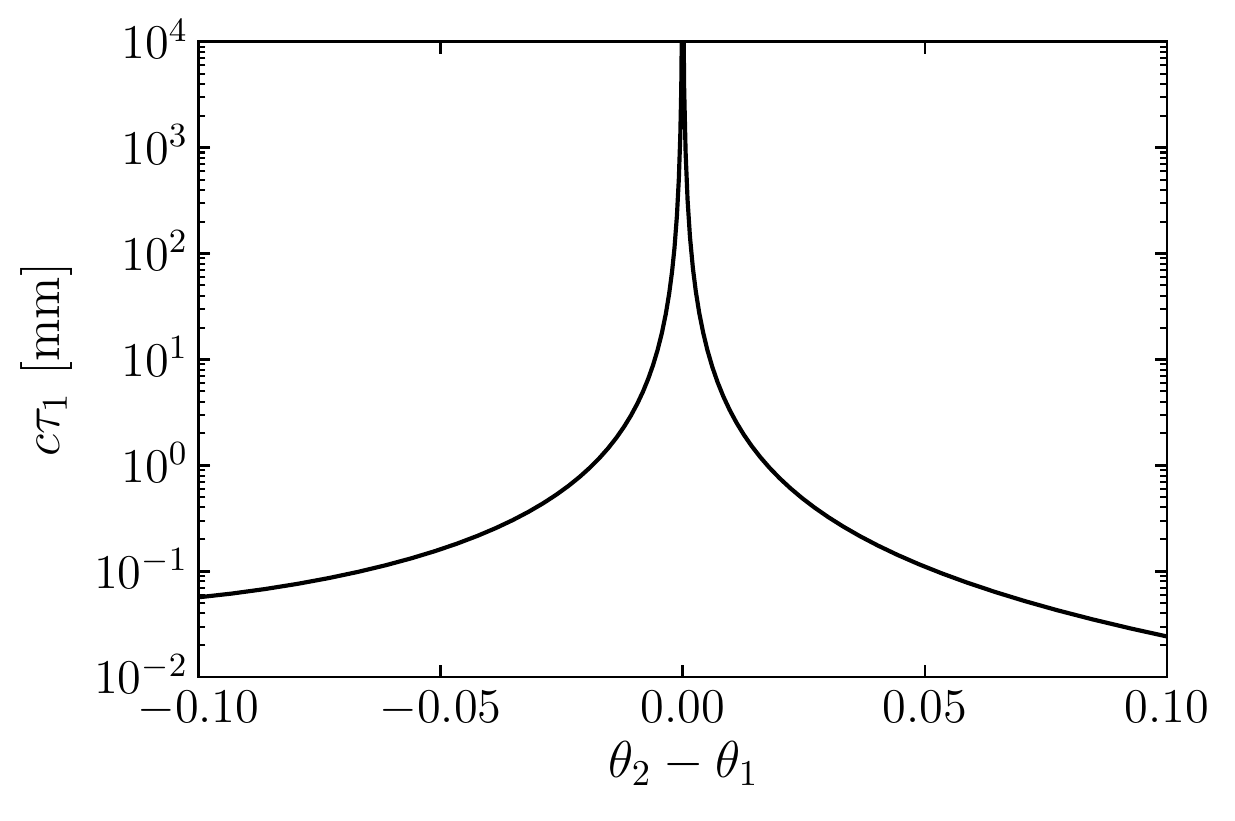}
    \caption{\it{The lightest scalar mass (on the left) and its flight distance (on the right) as a function of the $\mathcal{Z}_2$ breaking parameter associated to the $a_1$ particle. It is assumed that $\theta_1=1.1$.}}
    \label{fig:m+ctau}
\end{figure}

From here on, several parameters are fixed following the construction presented in Ref.~\cite{Blance:2019ixw}. Namely, $\gamma_1 \sim \gamma_2 \sim 1$ and $c_3 \sim - 4 (c_1 - 3 c_2)$, in order to turn off the mixing of the CP-even scalar with the Higgs boson and avoid the bounds from Higgs searches\footnote{Note that in other models leading to a smaller number of invariants, such as the one where ${{l_L} \oplus e_R = \mathbf{7} \oplus \mathbf{7}}$, the trilinear coupling $a_1^2 a_2$ is predicted to be of the same size as the parameter that controls the mixing with the Higgs particle.}.
The tadpole is removed by a suitable field redefinition. 
The trilinear scalar coupling, as well as the heaviest scalar mass, can be determined from the new physics scale and the two unknown UV couplings. Fixing $c_1 \sim  c_2  \sim g_*^2 /(4\pi)^2 \sim 10^{-6}$ by power counting~\cite{Giudice:2007fh}, with $g_* \sim 5$ and $f\gtrsim 1$ TeV, the heaviest scalar mass is predicted to be $m_2 \approx 3.1$ GeV while $m_{12} = 0.002$ GeV.
The mass dependence of the lightest scalar is represented in Figure~\ref{fig:m+ctau}, for $\theta_1\approx 1.1$. 

The Yukawa Lagrangian is also built upon three group invariants in the present model. The associated unknown constants in the UV are assumed to be of similar size, to prevent the generation of lepton flavour-changing neutral currents. Under this assumption
and keeping only the leading-order terms in $1/f$, the Yukawa Lagrangian reads:
\begin{equation}
L_{\rm yuk} = - y^\mu h \overline{\mu}_L\mu_R \bigg[1 + \frac{\left(1+\sqrt{2}\right) N }{f}a_2 + \ii \frac{\left(2+\sqrt{2}\right) N }{f} ( \theta_2 - \theta_1) a_1  \bigg]+\text{h.c.}\,,
\end{equation}
where $N= 2/(6+4\sqrt{2}-\theta_1^2) + \mathcal{O}(\theta_2-\theta_1) $.
In this work, the regime $\theta_1 \approx \theta_2$ is studied, so that the lightest scalar particle becomes long-lived at collider scales. (Such regime is the soft-breaking limit of a $\mathcal{Z}_2$ symmetry under which $a_1 \to - a_1$.) On the other hand, the heaviest scalar in this model is expected to be prompt, with $g_2\sim 0.1$. 
Furthermore, these coupling values are compatible with masses of $\mathcal{O}(1)$ GeV, as illustrated in Figure~\ref{fig:m+ctau}. 

\newpage

\section{Mass-dependent maps}~\label{sec:maps}
\begin{figure}[h!]
    \centering \includegraphics[width=0.49\textwidth]{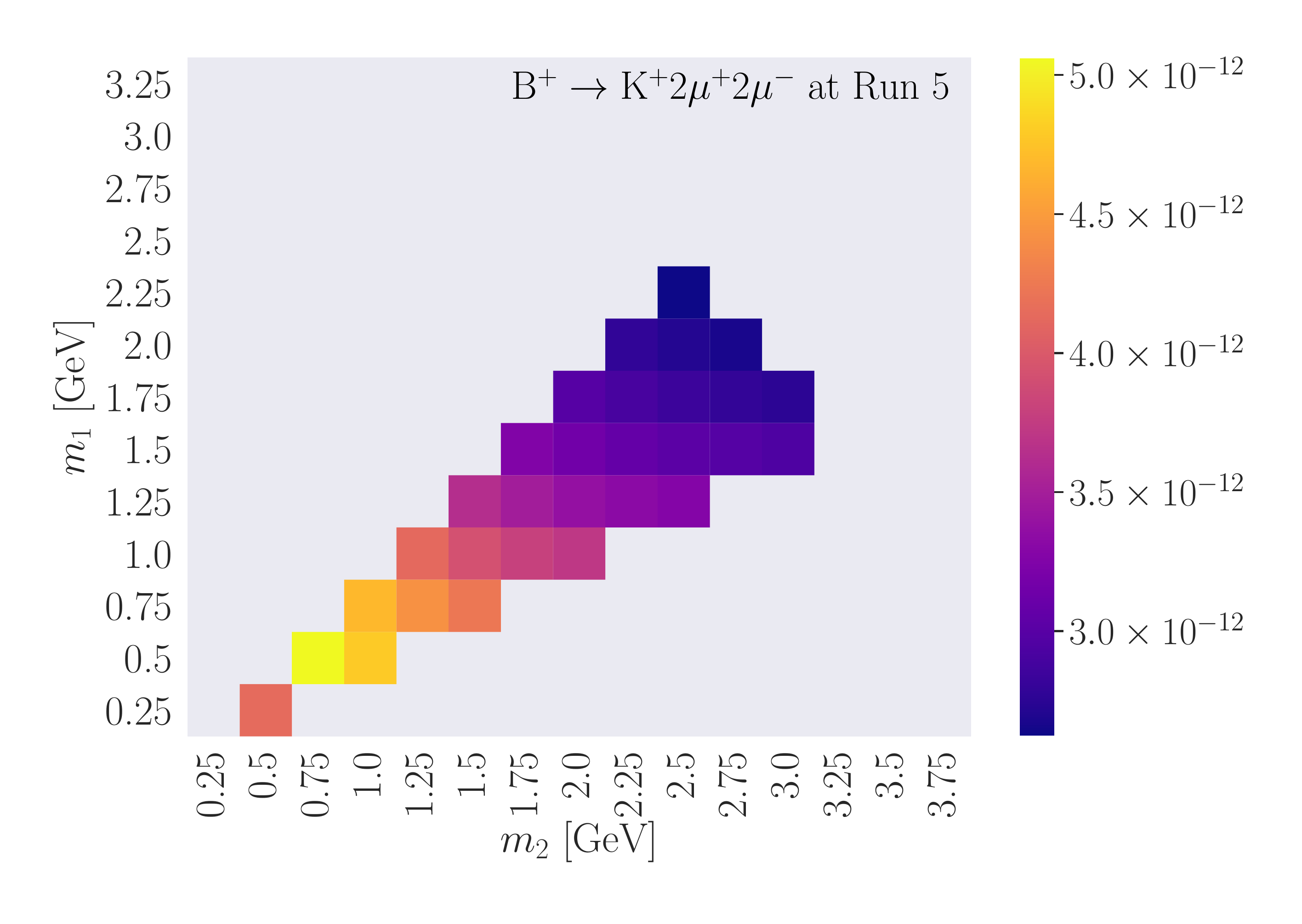}
    \includegraphics[width=0.49\textwidth]{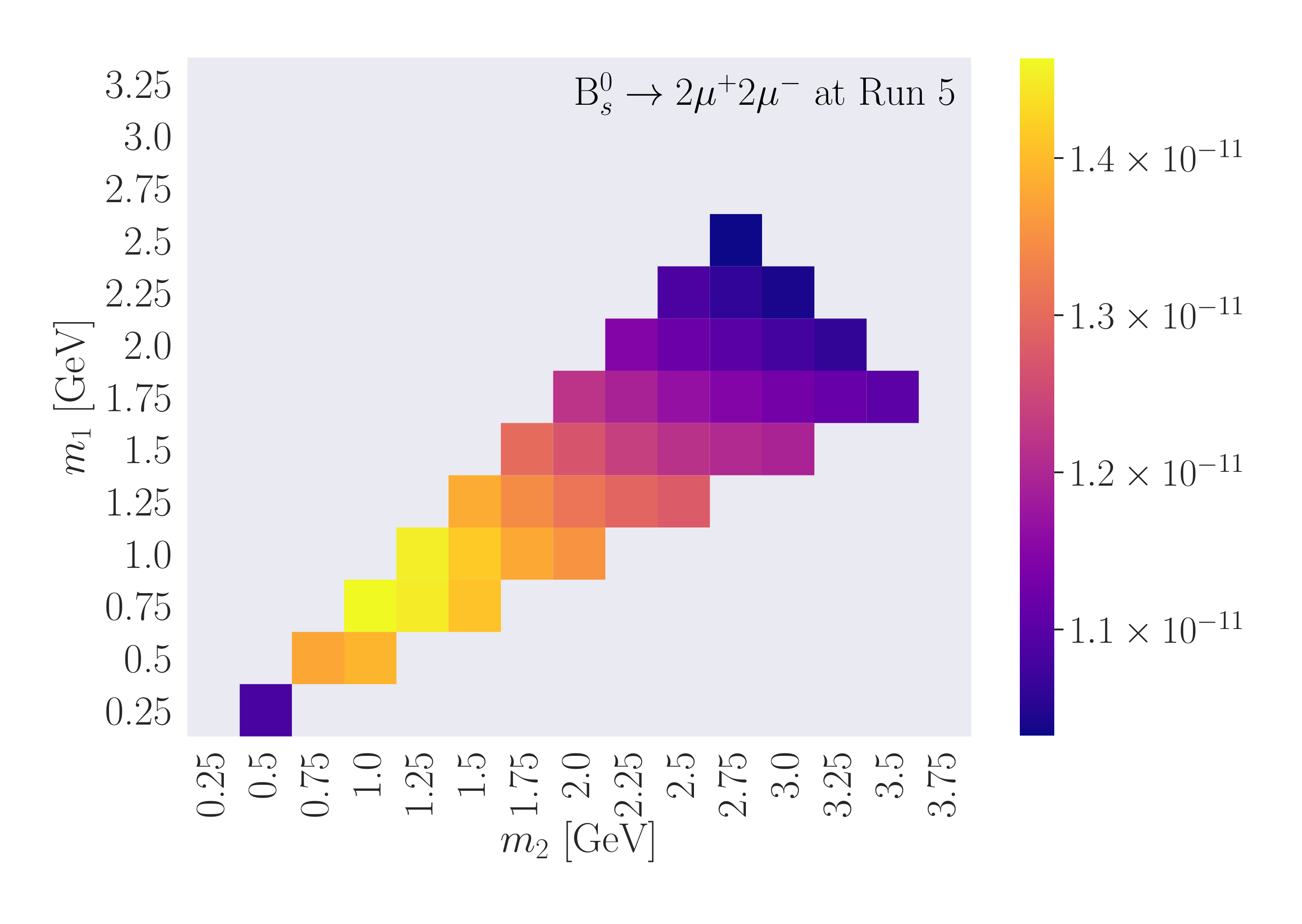}
    \newline
    \includegraphics[width=0.49\textwidth]{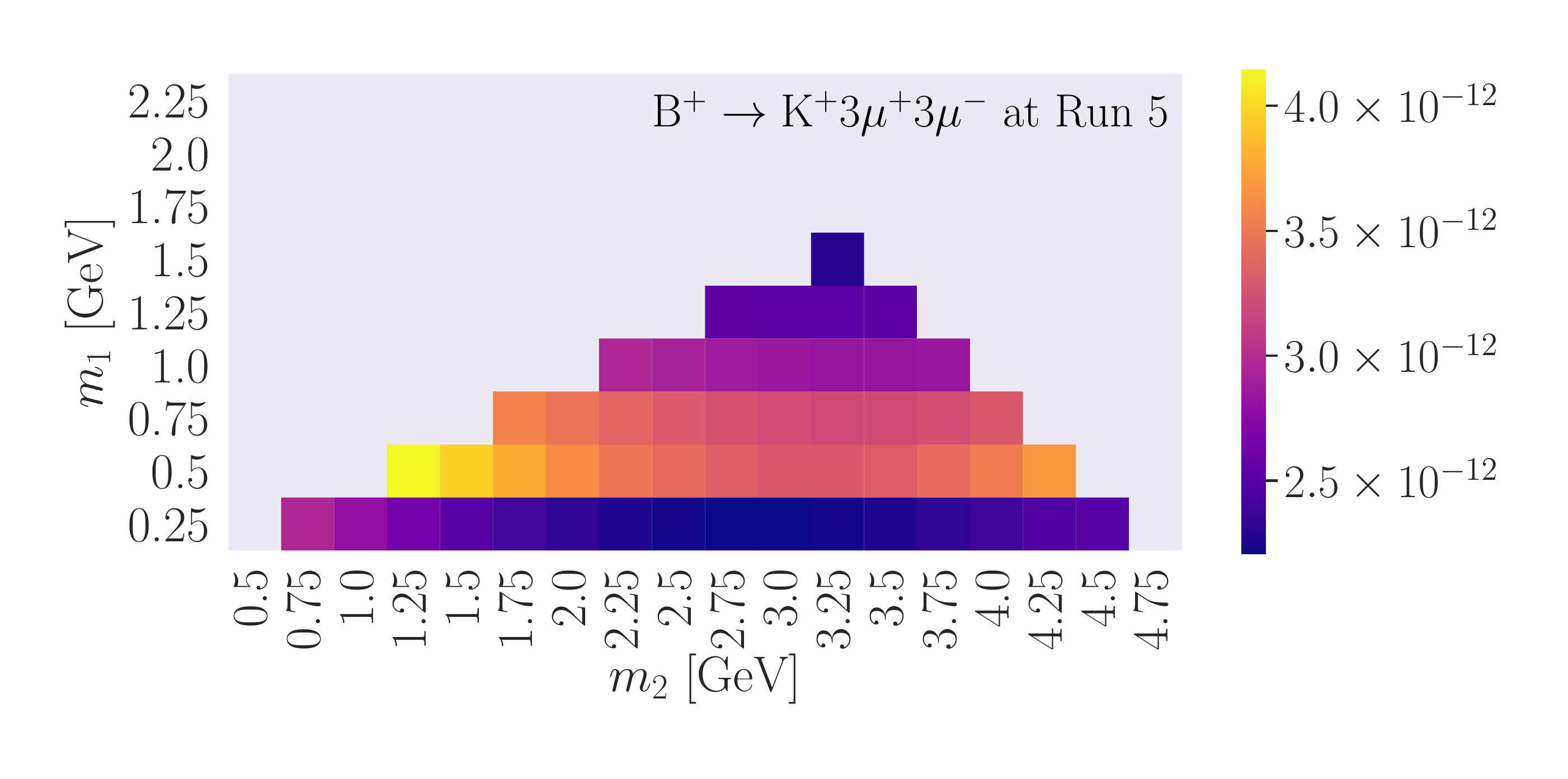}
    \includegraphics[width=0.49\textwidth]{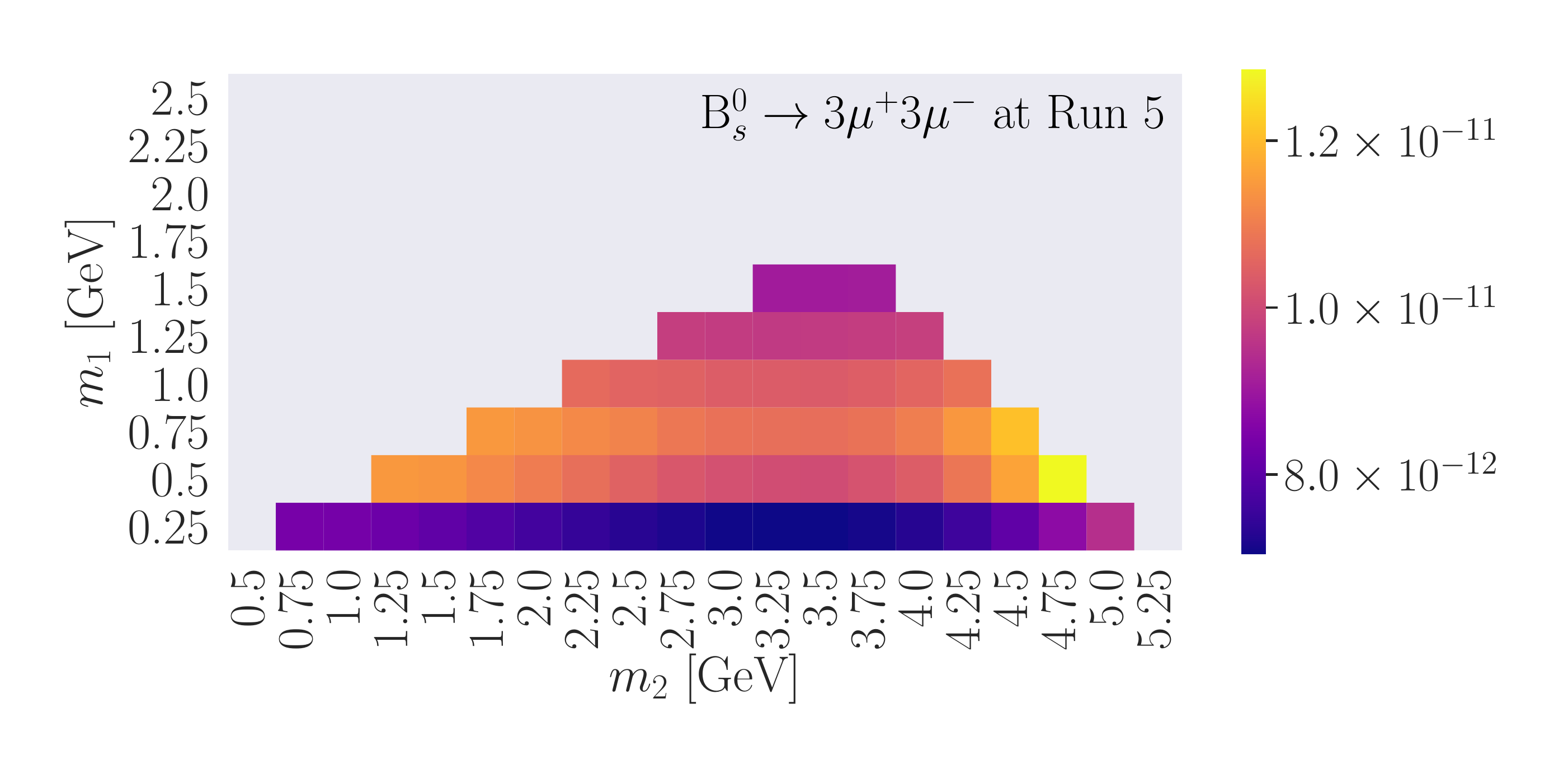}
    \label{fig:limwoK}
    \caption{\it Mass-dependent expected limits at Run 5 on the branching fractions of (top left) ${B^{+} \to K^{+} 2\mu^{+} 2\mu^{-}}$, (top right) ${B_{s}^{0} \to 2\mu^{+} 2\mu^{-}}$, (bottom left) ${B^{+} \to K^{+} 3\mu^{+} 3\mu^{-}}$ and (bottom right) ${B_{s}^{0} \to 3\mu^{+} 3\mu^{-}}$ for a lifetime of the long-lived scalar of $1\,\mathrm{ps}$.
    The top figures require that $m_{2} \leq 2 m_{1}$.
    The decays are reconstructed inclusively by requiring four muons as Long tracks.}
\end{figure}

\begin{figure}[h!]
    \centering \includegraphics[width=0.49\textwidth]{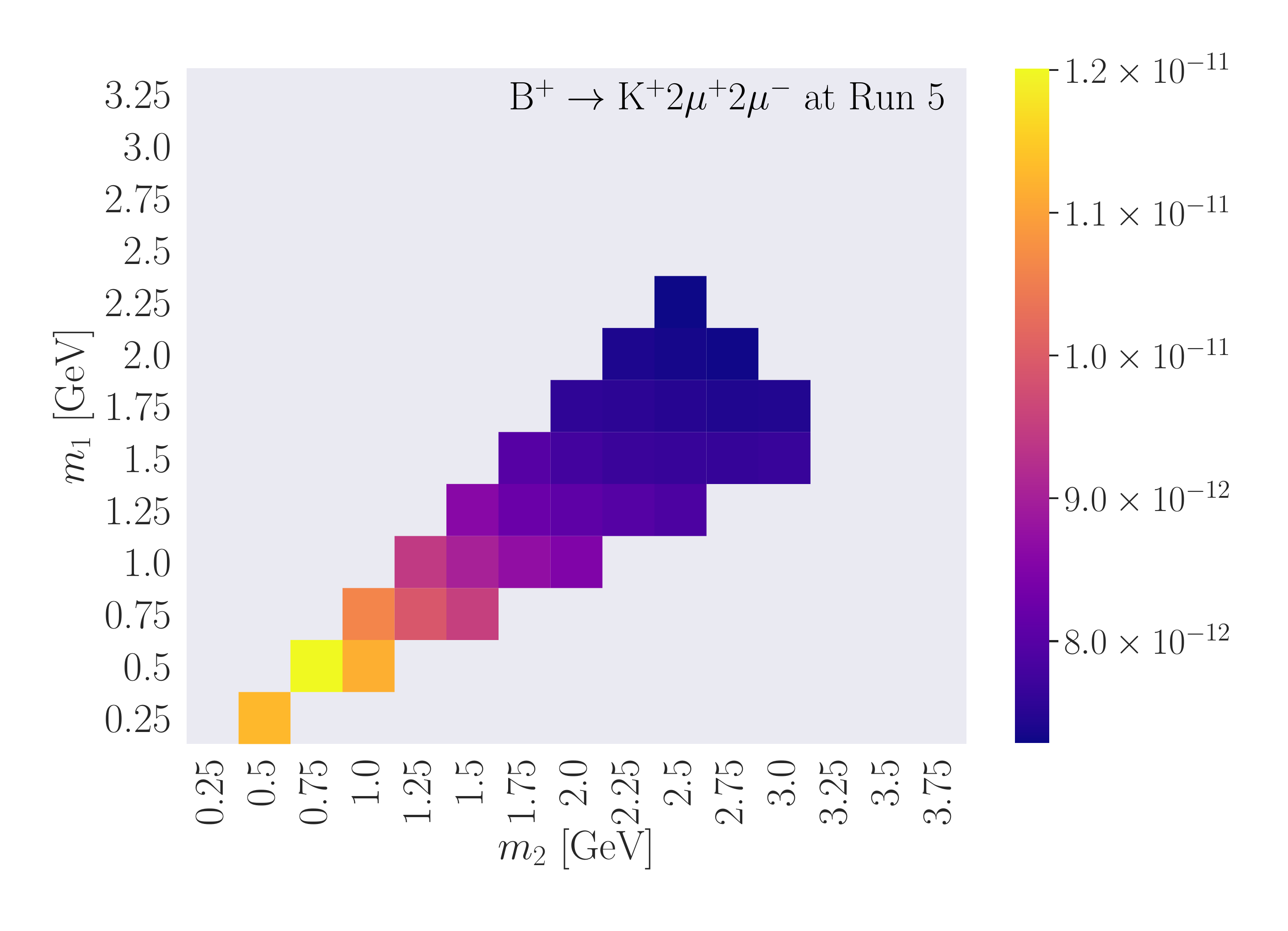}    \includegraphics[width=0.49\textwidth]{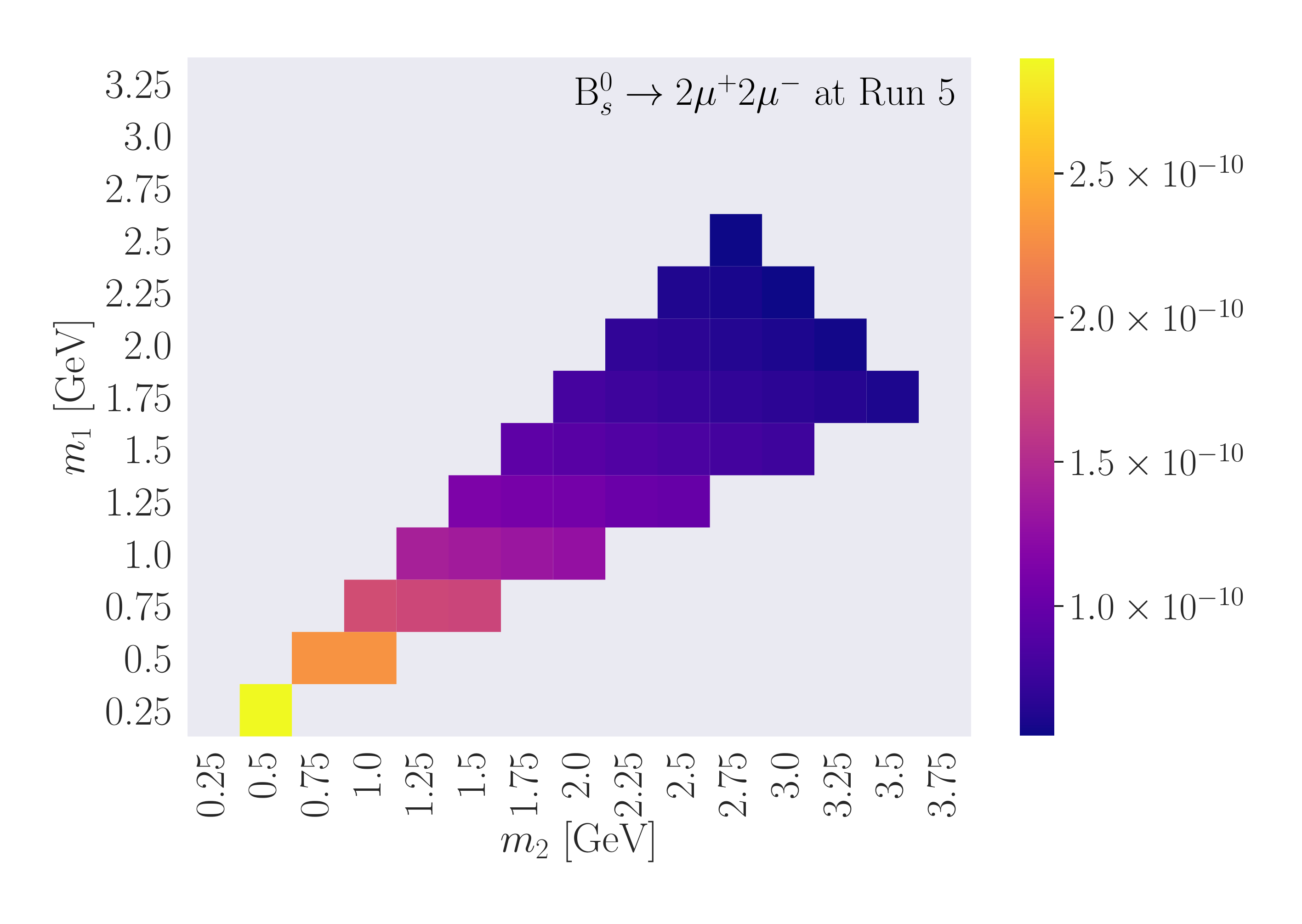}
    \newline
    \includegraphics[width=0.49\textwidth]{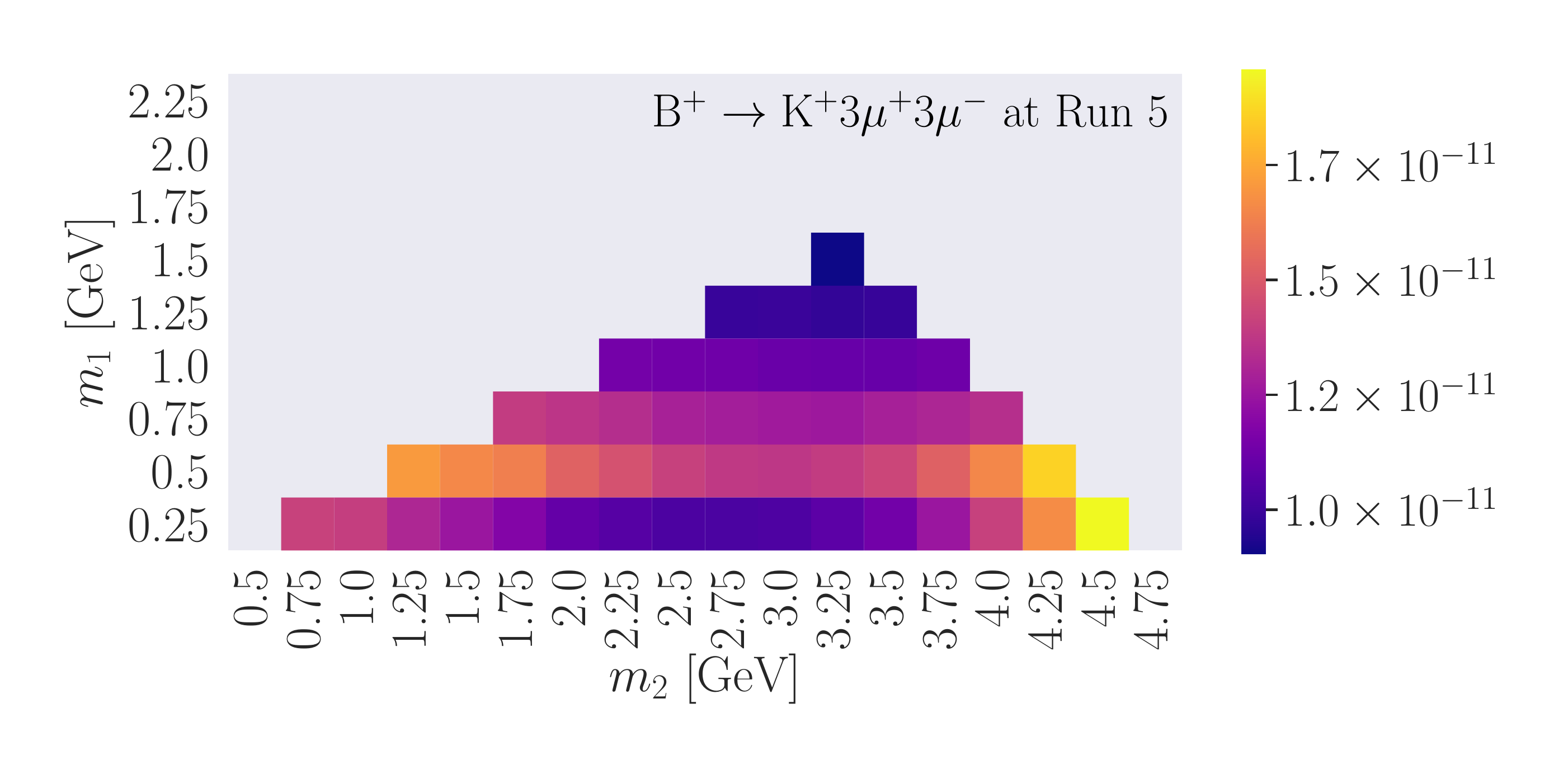}
    \includegraphics[width=0.49\textwidth]{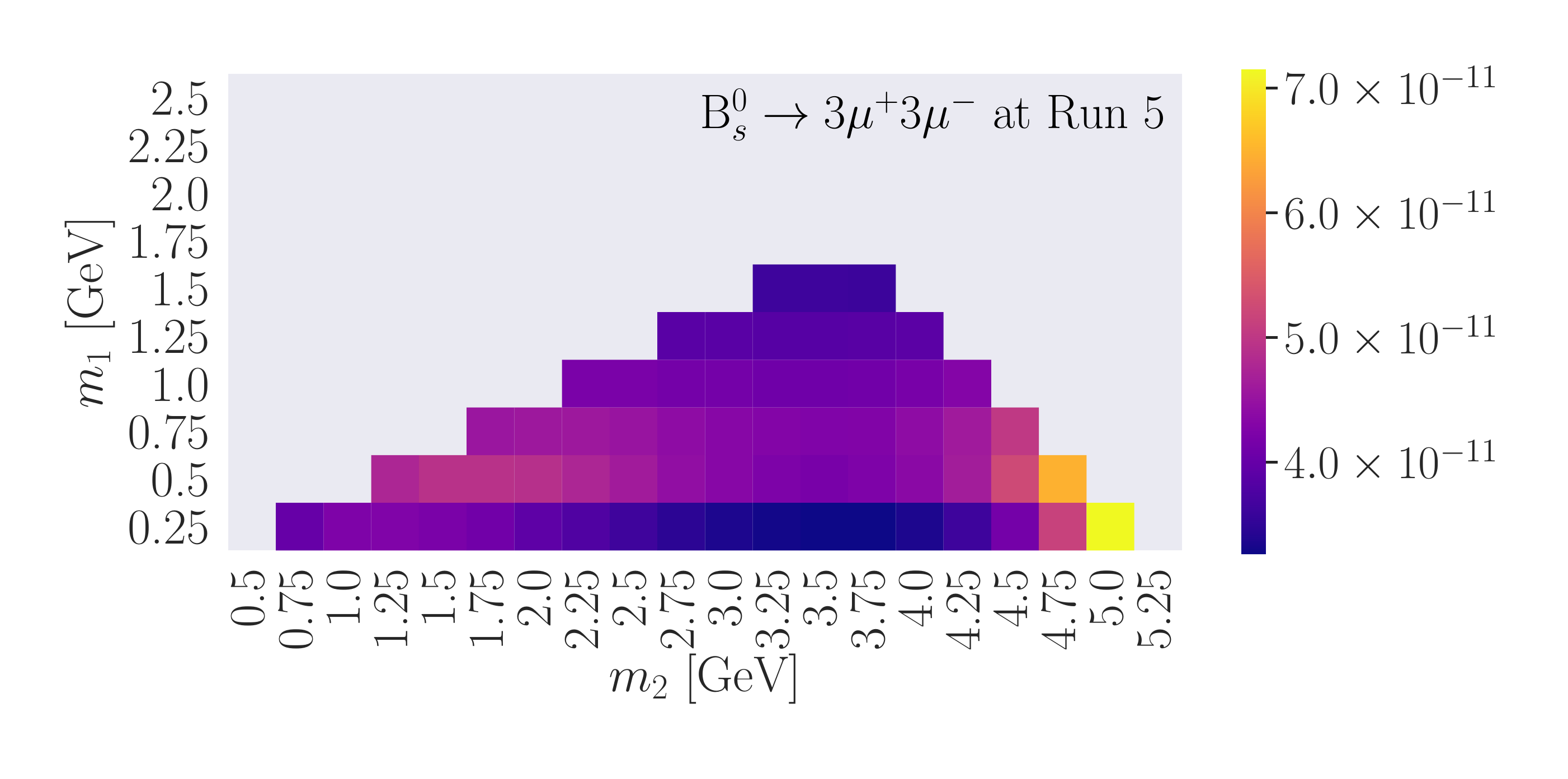}

    \caption{\it Mass-dependent expected limits at Run 5 on the branching fractions of (top left) ${B^{+} \to K^{+} 2\mu^{+} 2\mu^{-}}$, (top right) ${B_{s}^{0} \to 2\mu^{+} 2\mu^{-}}$, (bottom left) ${B^{+} \to K^{+} 3\mu^{+} 3\mu^{-}}$ and (bottom right) ${B_{s}^{0} \to 3\mu^{+} 3\mu^{-}}$ for a lifetime of the long-lived scalar of $100\,\mathrm{ps}$.
    The top figures require that $m_{2} \leq 2 m_{1}$.
    The decays are reconstructed inclusively by requiring four muons as Downstream tracks.}
    \label{fig:limwK}
\end{figure}

\begin{figure}[h!]
    \centering \includegraphics[width=0.49\textwidth]{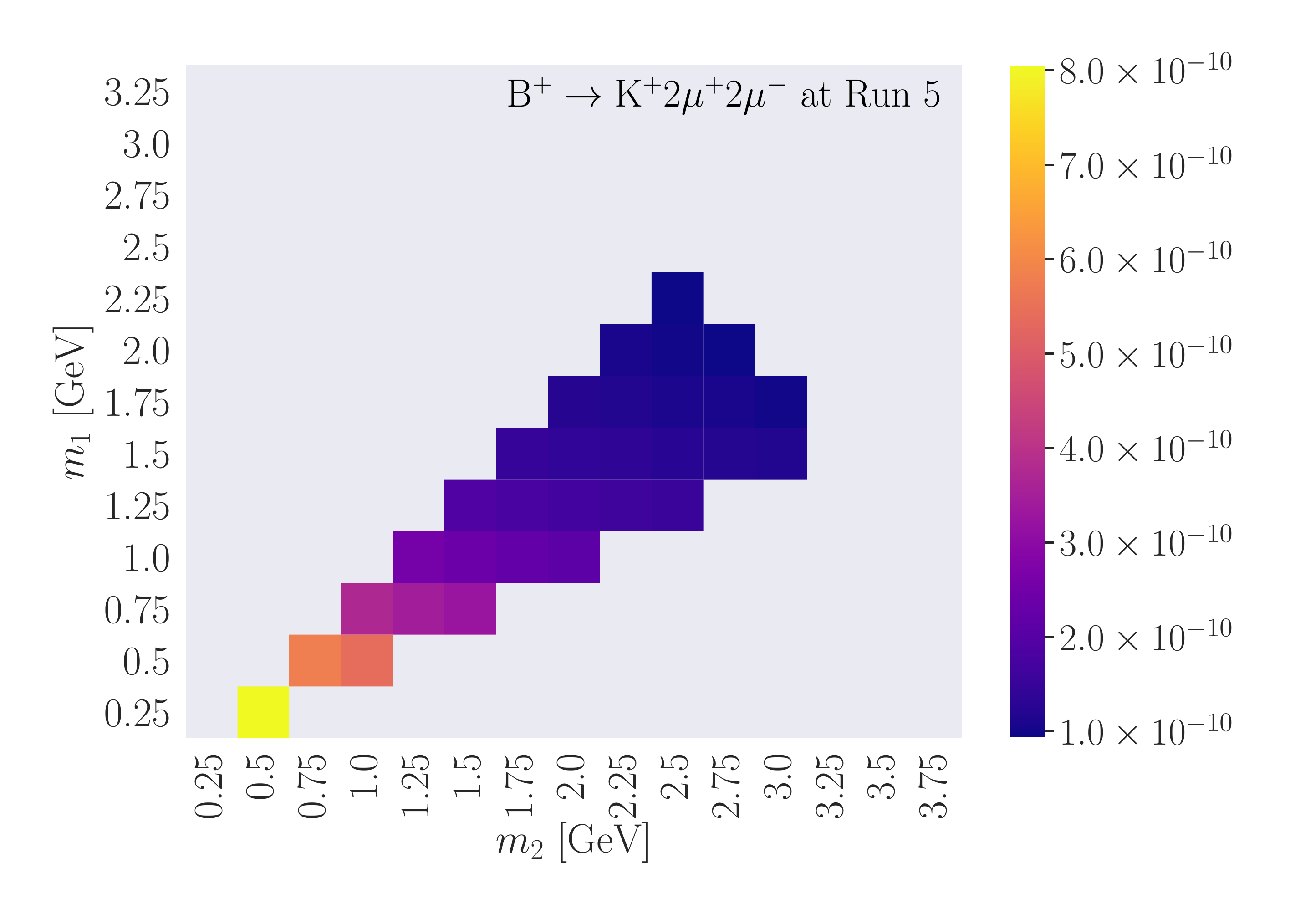}    \includegraphics[width=0.49\textwidth]{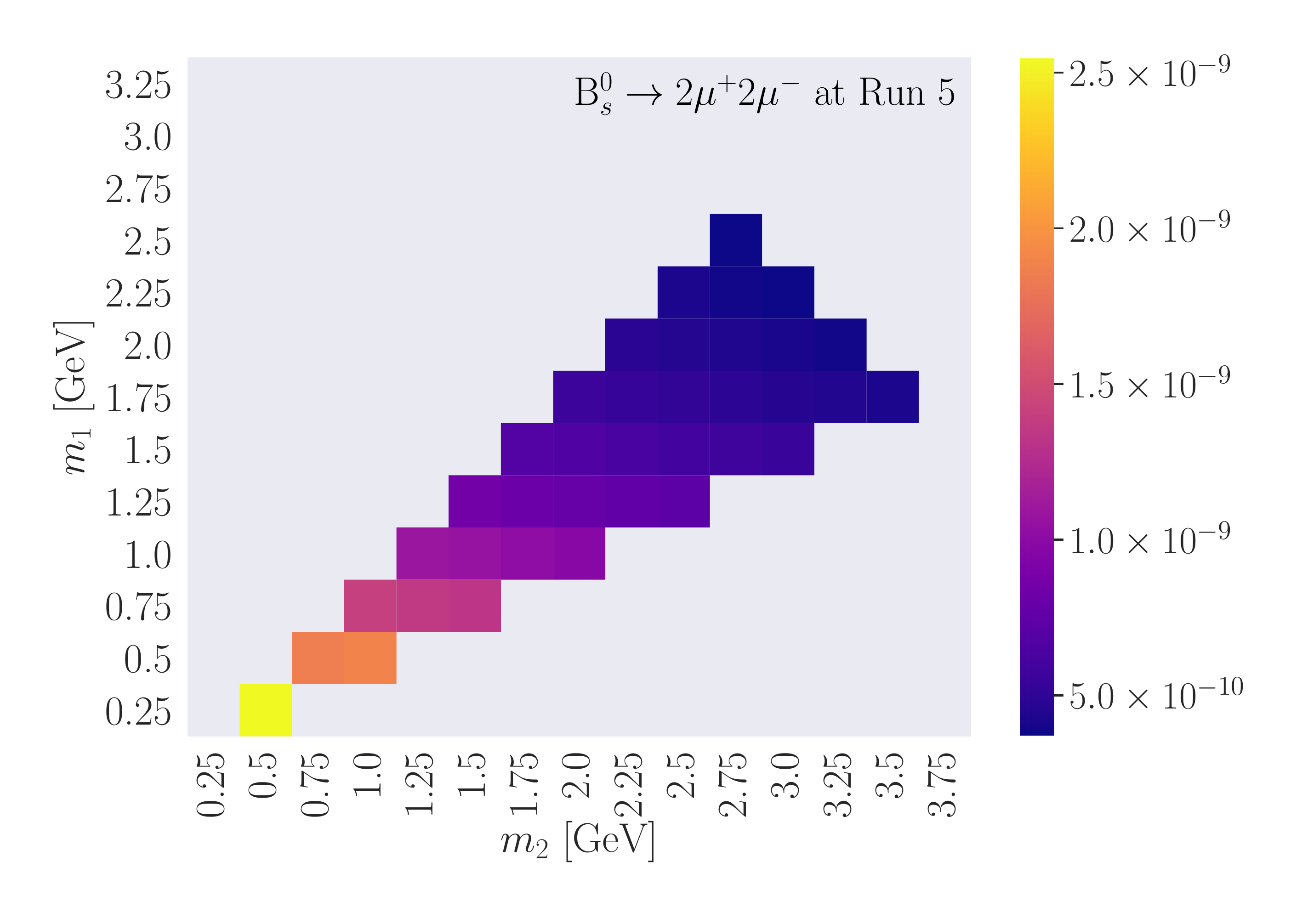}
    \newline
    \includegraphics[width=0.49\textwidth]{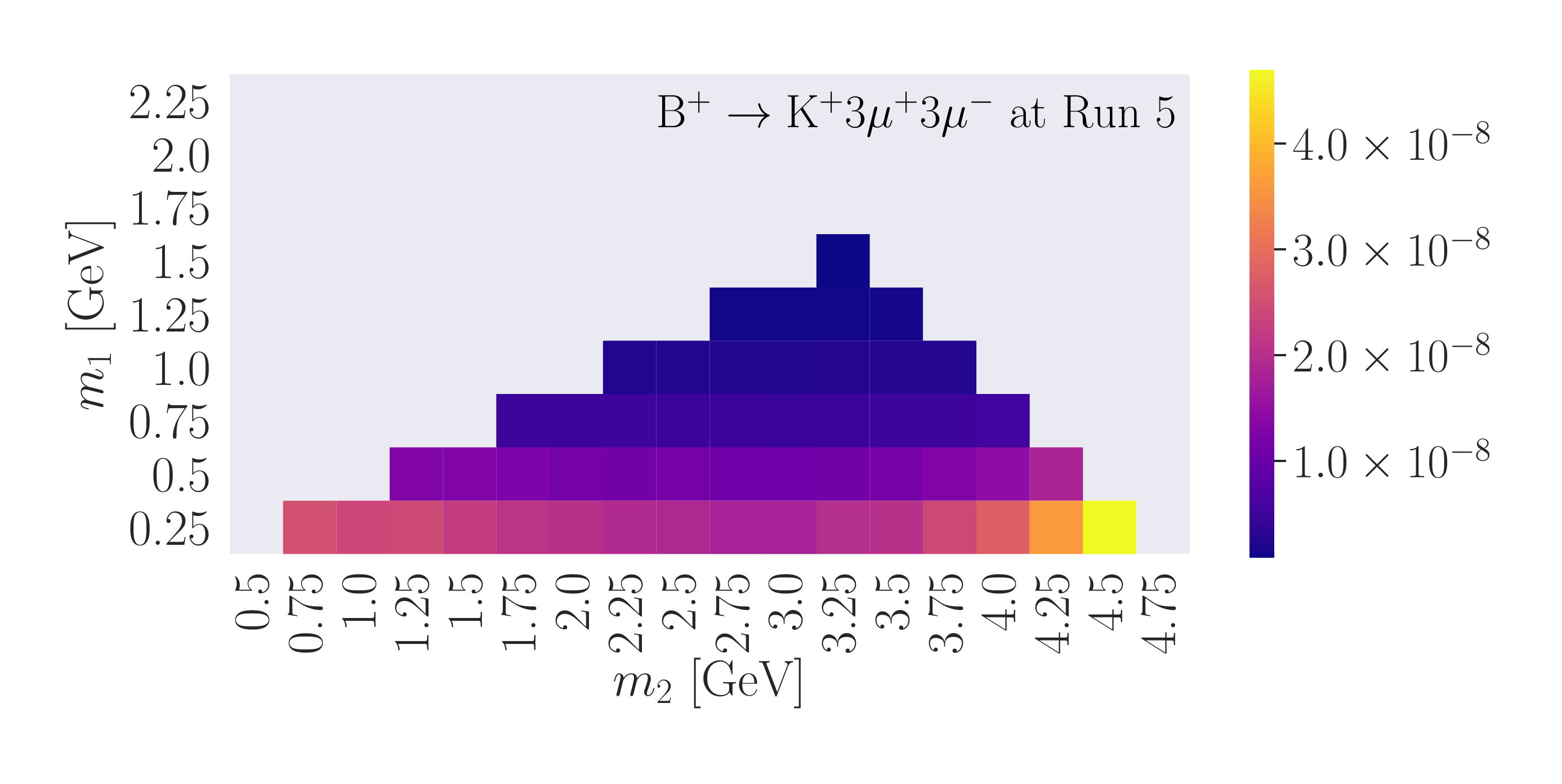}
    \includegraphics[width=0.49\textwidth]{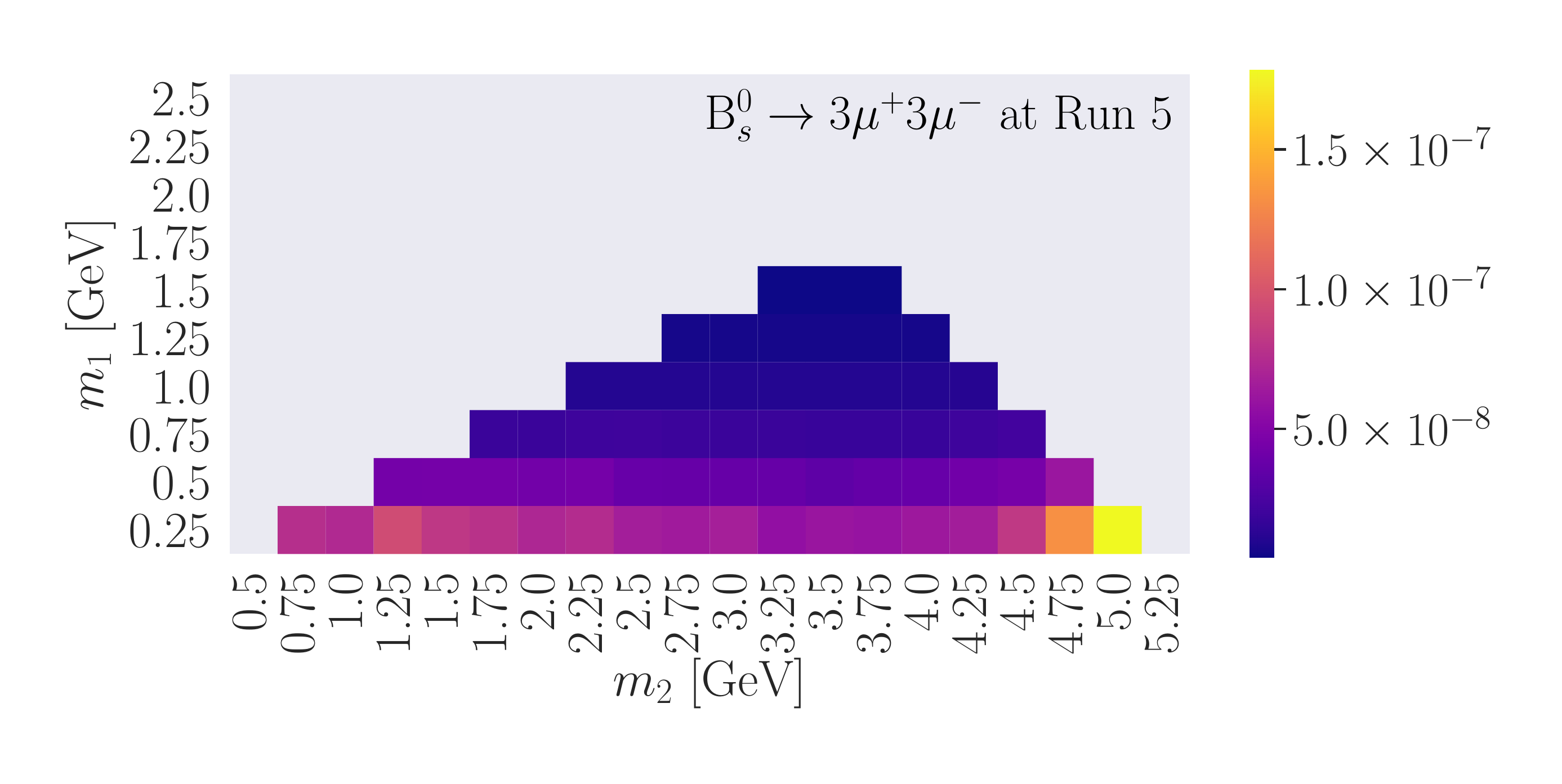}

    \caption{\it Mass-dependent expected limits at Run 5 on the branching fractions of (top left) ${B^{+} \to K^{+} 2\mu^{+} 2\mu^{-}}$, (top right) ${B_{s}^{0} \to 2\mu^{+} 2\mu^{-}}$, (bottom left) ${B^{+} \to K^{+} 3\mu^{+} 3\mu^{-}}$ and (bottom right) ${B_{s}^{0} \to 3\mu^{+} 3\mu^{-}}$ for a lifetime of the long-lived scalar of $10\,\mathrm{ns}$.
    The top figures require that $m_{2} \leq 2 m_{1}$.
    The decays are reconstructed inclusively by requiring four muons as T-tracks.}
    \label{fig:mass_distr_10Mfs}
\end{figure}

Figures~\ref{fig:limwoK},~\ref{fig:limwK} and ~\ref{fig:mass_distr_10Mfs} show the mass-dependent expected upper limit distributions across different lifetimes using the inclusive reconstruction method described in the paper. The sensitivities vary less than a factor of 2, with the strongest limits being obtained for extreme values of $m_1$.

\end{document}